\documentstyle[12pt,epsfig]{article}

\catcode`\@=11
\def\@citex[#1]#2{%
\if@filesw \immediate \write \@auxout {\string \citation {#2}}\fi
\@tempcntb\m@ne \let\@h@ld\relax \def\@citea{}%
\@cite{%
  \@for \@citeb:=#2\do {%
    \@ifundefined {b@\@citeb}%
      {\@h@ld\@citea\@tempcntb\m@ne{\bf ?}%
      \@warning {Citation `\@citeb ' on page \thepage \space undefined}}%
      {\@tempcnta\@tempcntb \advance\@tempcnta\@ne%
      \@tempcntb\number\csname b@\@citeb \endcsname \relax%
      \ifnum\@tempcnta=\@tempcntb 
	\ifx\@h@ld\relax%
	  \edef \@h@ld{\@citea\csname b@\@citeb\endcsname}%
	\else%
	  \edef\@h@ld{\ifmmode{-}\else--\fi\csname b@\@citeb\endcsname}%
	\fi%
      \else
	\@h@ld\@citea\csname b@\@citeb \endcsname%
	\let\@h@ld\relax%
      \fi}%
    \def\@citea{,\penalty\@highpenalty\,}%
  }\@h@ld
}{#1}}

\def\@citeb#1#2{{[#1]\if@tempswa , #2\fi}}
\def\@citeu#1#2{{$^{#1}$\if@tempswa , #2\fi }}
\def\@citep#1#2{{#1\if@tempswa , #2\fi}}

\def\bcites{         
	\catcode`\@=11
	\let\@cite=\@citeb
	\catcode`\@=12
}

\def\upcites{         
	\catcode`\@=11
	\let\@cite=\@citeu
	\catcode`\@=12
}

\def\plaincites{      
	\catcode`\@=11
	\let\@cite=\@citep
	\catcode`\@=12
}

\newcount\hour
\newcount\minute
\newtoks\amorpm
\hour=\time\divide\hour by 60
\minute=\time{\multiply\hour by 60 \global\advance\minute by-\hour}
\edef\standardtime{{\ifnum\hour<12 \global\amorpm={am}%
	\else\global\amorpm={pm}\advance\hour by-12 \fi
	\ifnum\hour=0 \hour=12 \fi
	\number\hour:\ifnum\minute<10 0\fi\number\minute\the\amorpm}}
\edef\militarytime{\number\hour:\ifnum\minute<10 0\fi\number\minute}

\def\draftlabel#1{{\@bsphack\if@filesw {\let\thepage\relax
   \xdef\@gtempa{\write\@auxout{\string
      \newlabel{#1}{{\@currentlabel}{\thepage}}}}}\@gtempa
   \if@nobreak \ifvmode\nobreak\fi\fi\fi\@esphack}
	\gdef\@eqnlabel{#1}}
\def\@eqnlabel{}
\def\@vacuum{}
\def\marginnote#1{}
\def\draftmarginnote#1{\marginpar{\raggedright\scriptsize\tt#1}}
\def\draft{
	\pagestyle{plain}
	\overfullrule=2pt
	\oddsidemargin -.5truein
	\def\@oddhead{\sl \phantom{\today\quad\militarytime} \hfil
	\smash{\Large\sl DRAFT} \hfil \today\quad\militarytime}
	\let\@evenhead\@oddhead
	\let\label=\draftlabel
	\let\marginnote=\draftmarginnote
	\def\ps@empty{\let\@mkboth\@gobbletwo
	\def\@oddfoot{\hfil \smash{\Large\sl DRAFT} \hfil}
	\let\@evenfoot\@oddhead}
	\def\@eqnnum{(\theequation)\rlap{\kern\marginparsep\tt\@eqnlabel}%
	\global\let\@eqnlabel\@vacuum}  }

\def\blackfonts{
	\font\blackboard=msbm10 scaled\magstep1
	\font\blackboards=msbm8
	\font\blackboardss=msbm6
}

\def\nblack{            
	\def\ZZ{{Z \n{10} Z}}
	\def\NN{{N \n{14} N}}
	\def\CC{{C \n{11} C}}
	\def\RR{{R \n{11} R}}
	\def\QQ{{Q \n{12} Q}}
	\def\PP{{P \n{11} P}}
}

\def\prep{         
	\catcode`\@=11
	\input art10.sty
	\catcode`\@=12
	
	\let\small\null
	\def\blackfonts{
		\font\blackboard=msbm10
		\font\blackboards=msbm7
		\font\blackboardss=msbm5
	}
	\let\sl\it
	\twocolumn
	\sloppy
	\voffset=-2.54truecm
	\hoffset=-2.54truecm
	\flushbottom
	\parindent 1em
	\leftmargini 2em
	\leftmarginv .5em
	\leftmarginvi .5em
	\marginparwidth 48pt
	\marginparsep 10pt
	\setlength{\columnsep}{2truecm}
	\setlength{\textwidth}{25.4truecm}
	\setlength{\textheight}{17truecm}
	\baselineskip=16pt
	\oddsidemargin .18truein
	\evensidemargin .17truein
}

\def\eqalign#1{\null\,\vcenter{\openup\jot\m@th
  \ialign{\strut\hfil$\displaystyle{##}$&$\displaystyle{{}##}$\hfil
      \crcr#1\crcr}}\,}
\def\eqalignno#1{\displ@y \tabskip\centering
  \halign to\displaywidth{\hfil$\@lign\displaystyle{##}$\tabskip\z@skip
    &$\@lign\displaystyle{{}##}$\hfil\tabskip\centering
    &\llap{$\@lign##$}\tabskip\z@skip\crcr
    #1\crcr}}

\def\section{\@startsection {section}{1}{\z@}{3.ex plus 1ex minus
 .2ex}{2.ex plus .2ex}{\large\bf}}
\def\subsection{\@startsection{subsection}{2}{\z@}{2.75ex plus 1ex minus
 .2ex}{1.5ex plus .2ex}{\bf}}

\def\appendix{{\newpage\section*{Appendix}}\let\appendix\section%
	{\setcounter{section}{0}
	\gdef\thesection{\Alph{section}}}\section}

\def\abstract{\if@twocolumn
\section*{Abstract}
\else 
\begin{center}
{\bf Abstract\vspace{-.5em}\vspace{0pt}}
\end{center}
\quotation
\fi}

\catcode`\@=12

\newcommand{\beq}{\begin{equation}}
\newcommand{\eeq}{\end{equation}}
\newcommand{\beqa}{\begin{eqnarray}}
\newcommand{\eeqa}{\end{eqnarray}}

\newcommand{\Z}{{\bf Z}}

\newcommand{\C}{{\bf C}}

\newcommand{\bk}{{\bf k}}
\newcommand{\bv}{{\bf v}}
\newcommand{\bw}{{\bf w}}
\newcommand{\bl}{{\bf l}}

\newcommand{\be}{\begin{equation}}
\newcommand{\ee}{\end{equation}}
\newcommand{\bea}{\begin{eqnarray}}
\newcommand{\eea}{\end{eqnarray}}

\def\noj#1,#2,{{\bf #1} (19#2)\ }
\def\jou#1,#2,#3,{{\sl #1\/ }{\bf #2} (19#3)\ }
\def\ann#1,#2,{{\sl Ann.\ Physics\/ }{\bf #1} (19#2)\ }
\def\cmp#1,#2,{{\sl Comm.\ Math.\ Phys.\/ }{\bf #1} (19#2)\ }
\def\ma#1,#2,{{\sl Math.\ Ann.\/ }{\bf #1} (19#2)\ }
\def\jd#1,#2,{{\sl J.\ Diff.\ Geom.\/ }{\bf #1} (19#2)\ }
\def\invm#1,#2,{{\sl Invent.\ Math.\/ }{\bf #1} (19#2)\ }
\def\cq#1,#2,{{\sl Class.\ Quantum Grav.\/ }{\bf #1} (19#2)\ }
\def\cqg#1,#2,{{\sl Class.\ Quantum Grav.\/ }{\bf #1} (19#2)\ }
\def\ijmp#1,#2,{{\sl Int.\ J.\ Mod.\ Phys.\/ }{\bf A#1} (19#2)\ }
\def\jmphy#1,#2,{{\sl J.\ Geom.\ Phys.\/ }{\bf #1} (19#2)\ }
\def\jams#1,#2,{{\sl J.\ Amer.\ Math.\ Soc.\/ }{\bf #1} (19#2)\ }
\def\grg#1,#2,{{\sl Gen.\ Rel.\ Grav.\/ }{\bf #1} (19#2)\ }
\def\mpl#1,#2,{{\sl Mod.\ Phys.\ Lett.\/ }{\bf A#1} (19#2)\ }
\def\nc#1,#2,{{\sl Nuovo Cim.\/ }{\bf #1} (19#2)\ }
\def\np#1,#2,{{\sl Nucl.\ Phys.\/ }{\bf B#1} (19#2)\ }
\def\pl#1,#2,{{\sl Phys.\ Lett.\/ }{\bf #1B} (19#2)\ }
\def\pla#1,#2,{{\sl Phys.\ Lett.\/ }{\bf #1A} (19#2)\ }
\def\pr#1,#2,{{\sl Phys.\ Rev.\/ }{\bf #1} (19#2)\ }
\def\prd#1,#2,{{\sl Phys.\ Rev.\/ }{\bf D#1} (19#2)\ }
\def\prl#1,#2,{{\sl Phys.\ Rev.\ Lett.\/ }{\bf #1} (19#2)\ }
\def\prp#1,#2,{{\sl Phys.\ Rept.\/ }{\bf #1C} (19#2)\ }
\def\ptp#1,#2,{{\sl Prog.\ Theor.\ Phys.\/ }{\bf #1} (19#2)\ }
\def\ptpsup#1,#2,{{\sl Prog.\ Theor.\ Phys.\/ Suppl.\/ }{\bf #1} (19#2)\ }
\def\rmp#1,#2,{{\sl Rev.\ Mod.\ Phys.\/ }{\bf #1} (19#2)\ }
\def\yadfiz#1,#2,#3[#4,#5]{{\sl Yad.\ Fiz.\/ }{\bf #1} (19#2) #3%
\ [{\sl Sov.\ J.\ Nucl.\ Phys.\/ }{\bf #4} (19#2) #5]}
\def\zh#1,#2,#3[#4,#5]{{\sl Zh.\ Exp.\ Theor.\ Fiz.\/ }{\bf #1} (19#2) #3%
\ [{\sl Sov.\ Phys.\ JETP\/ }{\bf #4} (19#2) #5]}

\def\beq{\begin{equation}}
\def\eeq{\end{equation}}

\newcommand{\bM}{\overline{\cal M}}
\newcommand{\M}{{\cal M}}

\def\nfrac#1#2{{\displaystyle{\vphantom1\smash{\lower.5ex\hbox{\small$#1$}}%
	\over\vphantom1\smash{\raise.25ex\hbox{\small$#2$}}}}}

\def\p#1{\mskip#1mu}
\def\n#1{\mskip-#1mu}
\def\stop{\p6.}
\def\comma{\p6,}

\def\to{\rightarrow}

\def\lae{\mathrel{\mathop{\smash{\lower .5 ex \hbox{$\stackrel<\sim$}}}}}
\def\lae{\mathrel{\mathop{\smash{\lower .5 ex \hbox{$\stackrel>\sim$}}}}}

\def\l:{\mathopen{:}\,}
\def\r:{\,\mathclose{:}}

\def\vm{\vec{m}}

\catcode`\@=11
\def\theequation{\arabic{equation}}
\catcode`\@=12

\nblack
\bcites

\nblack

\catcode`\@=11
\def\theequation{\thesection.\arabic{equation}}
\@addtoreset{equation}{section}
\@addtoreset{footnote}{section}
\@addtoreset{footnote}{subsection}
\catcode`\@=12

\newcommand{\beqn}{\begin{equation}}
\newcommand{\eeqn}{\end{equation}}
\newcommand{\beqnarray}{\begin{eqnarray}}
\newcommand{\eeqnarray}{\end{eqnarray}}
\newcommand {\bear} [1] {\begin {array} {#1}}
\newcommand {\ear} {\end {array}}

\newcommand {\beqarn} {\begin{eqnarray*}}
\newcommand {\eeqarn} {\end{eqnarray*}}

\begin{document}
\begin{titlepage}

\begin{center}
\today
\hfill LBNL-039707, UCB-PTH-96/58 \\
\hfill                  hep-th/9612131

\vskip 1.5 cm
{\large \bf  Mirror Symmetry in Three-Dimensional 
Gauge Theories, $SL(2,Z)$ and D-Brane Moduli Spaces
\\}
\vskip 1 cm 
{Jan de Boer, Kentaro Hori, Hirosi Ooguri, Yaron Oz and Zheng Yin}\\
\vskip 0.5cm
{\sl Department of Physics,
University of California at Berkeley\\
366 Le\thinspace Conte Hall, Berkeley, CA 94720-7300, U.S.A.\\
and\\
Theoretical Physics Group, Mail Stop 50A--5101\\
Ernest Orlando Lawrence Berkeley National Laboratory, 
Berkeley, CA 94720, U.S.A.\\}

\end{center}

\vskip 0.5 cm
\begin{abstract}
We  construct intersecting D-brane configurations 
that encode the  gauge groups and field content of 
dual $N=4$ supersymmetric gauge theories in three dimensions.
The duality which exchanges the Coulomb and Higgs branches and the Fayet-Iliopoulos
and mass parameters is derived from the $SL(2,Z)$ symmetry of the type IIB string.
Using the D-brane configurations
we construct explicitly this mirror map between 
the dual theories and study the instanton corrections in the 
D-brane worldvolume theory via open string instantons.
A general procedure to obtain mirror pairs is presented and illustrated.  
We encounter transitions among different field theories 
that correspond to smooth movements in the D-brane moduli space.  
We discuss the relation between the 
duality of the gauge theories and the level-rank duality of affine Lie algebras.
Examples of other dual theories 
are presented and explained via 
T-duality and extremal transitions in type II string compactifications.
Finally we discuss a second way to study
instanton corrections in the gauge theory,
by wrapping five-branes 
around six-cycles
in $M-$theory compactified on a Calabi-Yau 4-fold.

\end{abstract}

\end{titlepage}

\section{Introduction}
Recently  a duality between $N=4$ supersymmetric gauge theories 
in three dimensions has been proposed under which the Higgs and Coulomb branches
and the Fayet-Iliopoulos (FI) and mass parameters are exchanged \cite{si}.
The dual gauge theories have an ALE space as Higgs branch, and were
based on Kronheimer's construction \cite{k}
of ALE spaces as a hyperk\"ahler quotient.
This duality has been generalized in \cite{dhoo}
to gauge theories whose Higgs branch is a
quiver variety. The gauge groups and field content
of the gauge theories is encoded in the quiver
diagrams that serve as the 
starting point for Kronheimer-Nakajima's
hyperk\"ahler quotient construction of
the quiver varieties \cite{kn,Na1}.
Various interpretations of the duality have been proposed in \cite{pz,hw,gomez}.

It was suggested in \cite{hw} that the duality can be 
interpreted as arising from the
$SL(2,Z)$ symmetry of type IIB string theory.
One of the aims of this paper is to 
apply this idea to the  
families of dual theories (called A and B models)
introduced and analyzed in \cite{dhoo}.

\noindent 
(1) The A-model has $U(k)$ gauge group, $n$ hypermultiplets
in the fundamental representation of the gauge group 
and one hypermultiplet in the adjoint representation. Its dual B-model 
has $U(k)^n$ gauge group and matter content specified
by a quiver diagram corresponding to the Hilbert scheme of $k$ points on an ALE space
of $A_{n-1}$ type.
By the Hilbert scheme of $k$ points on a complex surface $X$ we mean 
a smooth resolution
of the $k$-symmetric product of $X$, $Sym^k X$. Concretely,
there will be one hypermultiplet in the fundamental representation of one of the 
$U(k)$'s, and $n$ hypermultiplets 
charged under a pair of $U(k)$'s.

\noindent
(2) The A and B models have $U(k)^n$ and $U(k)^m$ gauge groups respectively,
and matter content specified
by quiver diagrams corresponding to the hyperk\"ahler quotient construction of
certain moduli spaces of instantons on vector bundles over  ALE spaces
of $A_{n-1}$ and $A_{m-1}$ types. All matter is charged under either one or
two $U(k)$ gauge groups. 

The paper is organized as follows:
In section 2 we associate intersecting D-brane configurations of type IIB string
with the quiver diagrams and
use the $SL(2,Z)$ symmetry
to construct their duals and to
derive the
mirror map between the mass terms of the A-model
and the FI terms of the  B-model, in agreement with \cite{dhoo}. 
The instanton corrections to the metric on the moduli space have an important role, 
as discussed in \cite{sw,dhoo,ch}.
We study them in the framework of intersecting D-brane configurations
as arising from open string instantons, and find 
agreement with what we expect from the field theory
analysis.  This enables us to gain further insight 
into the interelation between D-brane   
and field theory moduli spaces.  
In section 3 we study the condition for complete Higgsing in the
gauge theory corresponding to a general quiver
diagram using D-brane configurations. We rederive results which were proposed in
\cite{dhoo} from field theory viewpoint, and that were proven in \cite{Na2}.
The analysis of complete Higgsing provides
 a general procedure to obtain mirror pairs from 
quiver diagrams, which is illustrated by
 examples. We show that moving two 5-branes of 
 the same type through each other is reflected 
 as a phase transition on  
D3-brane worldvolume theory. 
We  then take another point of view and discuss 
 the gauge theory duality in relation to
 the level-rank duality of affine Lie algebras, which
are represented on the middle
homology of the moduli spaces of the gauge theories.
In section 4 we  consider dual Abelian gauge theories.
We prove the duality  using  field theory
methods as well as 
T-duality and extremal transitions in type II string compactifications. 
Finally, in section 5
we study once more the instanton corrections, this time
by relating them to the wrappings of
five-branes in $M-$theory around divisors of Calabi-Yau 4-folds, in
a similar fashion as in  \cite{witten}.

\section{Quivers and Intersecting Branes}

In this section we construct configurations of intersecting 
three and five-branes in type IIB
string theory, in such a way that the world-volume theory
on the three-branes  has the gauge groups and matter content associated with
quiver diagrams.
We will then use the $SL(2,Z)$ symmetry of type IIB string theory to study mirror symmetry
and verify part of the results of \cite{dhoo}.

Following \cite{hw} we use NS 5-branes, Dirichlet 5-branes  and Dirichlet 3-branes 
in order to construct configurations that preserve one quarter of the space-time supersymmetry.
We will use conventions and notations similar to those used in \cite{hw}:
The worldvolume coordinates of the NS 5-branes, the Dirichlet 5-branes and
the Dirichlet 3-branes are $(x^0,x^1,x^2,x^3,x^4,x^5)$, 
$(x^0,x^1,x^2,x^7,x^8,x^9)$ and $(x^0,x^1,x^2,x^6)$ respectively.
The coordinate $x^6$, which is one of the dimensions of the worldvolume of the Dirichlet 3-branes
 is compactified on a circle of radius $R$.
The fact that  the coordinate $x^6$ is compactified on a circle
 will change the field content of the worldvolume theory compared to
the cases studied in \cite{hw} where the coordinate $x^6$ took values on the
real line.

The position of the $i$th NS 5-brane in  $(x^7,x^8,x^9)$  will be denoted by $\vec{\omega}_i$.
Between the $(i-1)$th and $i$th NS 5-brane there will be $k_i$ 
3-branes, whose world-volume theory contains a $U(k_i)$ gauge group. 
The Fayet-Iliopoulos parameters 
$\{\vec{\zeta}_{i}\}$
for this $U(k_i)$ gauge group are related to the 
parameters $\{\vec{\omega}_i\}$
by
\beq
 \vec{\zeta}_{i} = \vec{\omega}_{i} - \vec{\omega}_{i-1}
 \stop
 \label{zeta}
 \eeq
The position of the $i$th  Dirichlet 5-brane in  $(x^3,x^4,x^5)$  will be denoted by $\vm_i$
and will correspond to the  mass parameter associated with the  $i$th  hypermultiplet in the
fundamental representation.
This hypermultiplet arises from an open string stretching between the 
$i$th Dirichlet 5-brane and the Dirichlet 3-brane.
The position of the  Dirichlet 3-brane in  $(x^3,x^4,x^5)$ 
will correspond to the  vev's of the scalars in the vector multiplet 
which together with the vev's of the 
scalars dual to the vector fields on the  Dirichlet 3-brane 
worldvolume parameterize 
the vector multiplet moduli space.
The position of the  Dirichlet 3-brane in  $(x^7,x^8,x^9)$  will be
non-linearly related 
to the vev's of the scalars in the hypermultiplets and  constitute  
part of the coordinates on the hypermultiplet moduli space. The  $x^6$
component of the vector field provides the remaining coordinates.
The $U(1)$ isometries of the hypermultiplet moduli space correspond to gauge transformations
of this component.

After compactification in the $x^6$-direction,
the three dimensional theory on the $(x^0,x^1,x^2)$ worldvolume of the Dirichlet 3-brane
is an $N=4$ supersymmetric gauge theory. From now on we will refer to this three dimensions
as the Dirichlet 3-brane worldvolume.
 The gauge 
coupling of the Dirichlet 3-brane
worldvolume theory is determined by the separation $r$ between the NS 5-branes on the circle.
In particular with one NS 5-brane we have the classical relation between the three and four dimensional
coupling constants

\beq
\frac{1}{g^2_3} = \frac{r}{g_4^2}
\label{R}
\eeq
where $r$ denotes, in this case, 
the radius of the compact $6$th direction. 
One expects, however, corrections to this classical formula.
The R-symmetry group is
$SU(2)_L\times SU(2)_R$  under which
the masses and FI parameters transform 
as $({\bf 3,1})$ and $({\bf 1,3})$ respectively. The mass parameters deform the metric
 on the Coulomb branch and lift some of the Higgs branch,
  while the FI parameters deform the metric on the Higgs branch and
  lift some of the Coulomb branch.
 The Higgs branch is constructed as a hyperk\"ahler quotient 
 with an $SU(2)_R$ action
 and is not modified
 by quantum corrections.

  Due to the $N=4$ supersymmetry the Coulomb branch is  a  hyperk\"ahler manifold with
 an $SU(2)_L$ action.
 Its metric is corrected by loop and monopole corrections.
 The monopoles are instantons in three dimensions and they provide exponential corrections
 to the metric.   
 
 The duality between $N=4$ supersymmetric gauge theories 
in three dimensions exchanges the  Higgs and Coulomb branches,
 the Fayet-Iliopoulos (FI) parameters and masses and the R-symmetry groups $SU(2)_L$ and
 $SU(2)_R$.

\subsection{Duality for $U(k)$ Gauge Groups}

Consider the intersecting D-brane configuration in figure 1a.

\begin{figure}
\begin{center}
$\mbox{\epsfig{figure=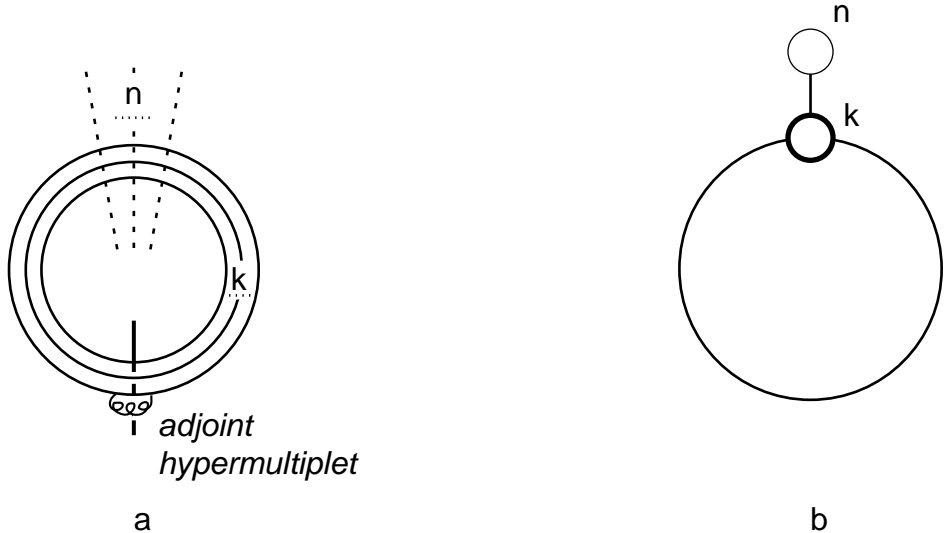}}$
\end{center}
\caption{The D-brane configuration of the A-model is plotted on the left.
The circles consist of $k$ Dirichlet 3-branes, the $n$ dashed lines are Dirichlet 5-branes
and the solid line is an NS 5-brane.
The
 corresponding quiver diagram is plotted on the right.}
\end{figure}
It consists of a NS 5-brane, $n$ Dirichlet 5-branes and $k$ Dirichlet 3-branes.
In order to read off the gauge group and matter content of
the three dimensional Dirichlet 3-brane worldvolume theory we have to apply the rules of \cite{hw}.
When $k$ Dirichlet 3-branes end on two NS 5-branes the gauge group is $U(k)$.
Since the NS 5-brane is positioned on a circle in figure 1a,
 the Dirichlet 3-branes do not have to end on it but can also be
 viewed
 as intersecting it and there is
 in addition a hypermultiplet in the adjoint representation arising from an open string
 connecting Dirichlet 3-branes
as depicted in figure 1a.
 In the absence of Dirichlet 5-branes, the adjoint hypermultiplet 
 together with the vector multiplet provide the field content for an $N=8$ supersymmetry on the 
 world volume of the Dirichlet 3-brane, which is the reduction to three dimensions
  of $N=1$ super Yang Mills in ten dimensions. 
 There are also $n$ hypermultiplets in the fundamental representation
 arising from the open  
strings connecting the $n$ Dirichlet 5-branes to the Dirichlet 3-branes.

The Dirichlet 3-branes worldvolume theory in $(x^0,x^1,x^2)$
is an
 $N=4$ supersymmetric theory 
with $U(k)$ gauge group,
$n$  hypermultiplets in the fundamental representation and one adjoint hypermultiplet.
We will use the terminology of \cite{dhoo}
and call this theory
the A-model.

The gauge group and matter content of the A-model
is associated with the quiver diagram in figure 1b, in a way which will be described below.
The field content is precisely what is needed for the 
hyperk\"ahler quotient
construction of the moduli space of $SU(n)$ 
$k-$instantons $\bM_k(SU(n))$ 
\cite{Donaldson},
which is 
the Higgs branch of the theory. 

We now perform an $SL(2,Z)$ transformation \footnote{In an $SL(2,Z)$ transformation we 
include  a rotation that exchanges the coordinates $(x^3,x^4,x^5)$ and
$(x^7,x^8,x^9)$.}  on the configuration of figure 1a.
Recall that under this transformation an NS 5-brane is transformed into
a Dirichlet 5-brane and vice versa, while the Dirichlet 3-brane is invariant. 
The $SL(2,Z)$ transformation of figure 1a
 yields  the configuration of 
of figure 2a, and the resulting gauge theory
corresponds exactly the quiver diagram of figure 2b.
This will be called the B-model.

\begin{figure}
\begin{center}
$\mbox{\epsfig{figure=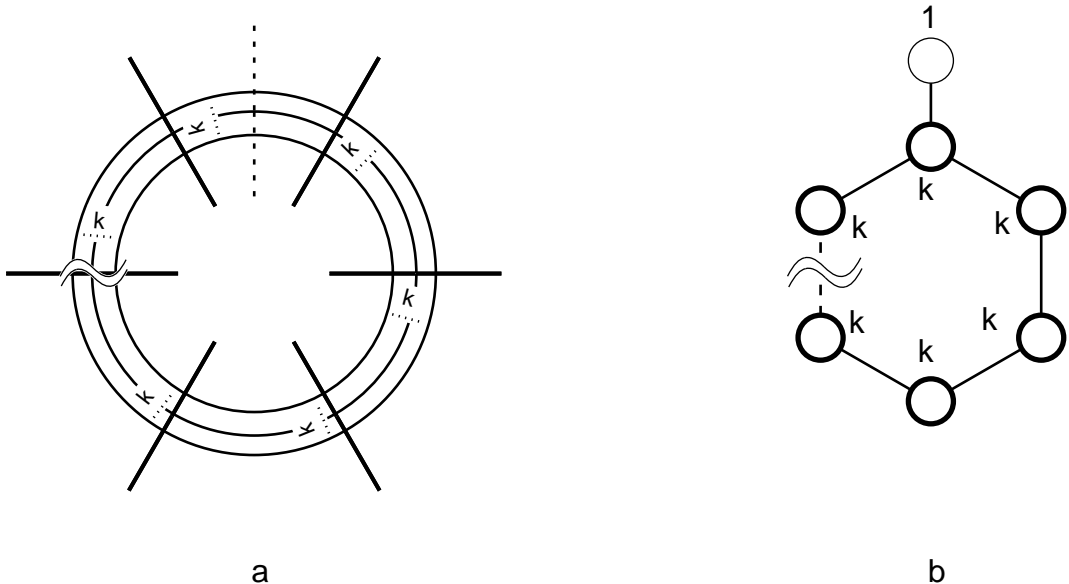}}$
\end{center}
\caption{The D-brane configuration of the B-model is plotted on the left, 
and the corresponding quiver diagram is plotted on the right.}
\end{figure}

The gauge group and matter content is encoded in the quiver diagrams in the following way.
Consider the quiver diagram in figure 2b, which contains figure 1a as a special case.
We attach an index $k_i$ at each node $i$. There
are $n$ nodes in the diagram with $k_i=k$ (indicated by thick circles)
and one node (indicated by a thin circle) with index $1$.
The  gauge group and the field content of the theory are 
encoded in the diagram in
the following way:
We associate to each node (indicated by thick circles)
 $i$ with $k_i=k$
a gauge group $U(k)_i$,
to each link (connecting two thick circles)
${}_i\!\circ\!\!-\!\!\!-\!\!\!-\!\circ_j$
with $k_i=k_j=k$ a hypermultiplet in the
representation $({\bf k},{\bf k^*})$ of $U(k)_i\times 
U(k)_j$,
and to the link (connecting a thick and a thin circle)
attached to the node with index 1
a hypermultiplet in the fundamental representation of 
the $U(k)$
gauge group associated with the other node of the link.
This is the field content needed for the hyperk\"ahler 
quotient
construction of the
Hilbert scheme of $k$ points on an ALE space
of type $A_{n-1}$, 
$X_{A_{n-1}}$\cite{kn,Na1},
 which is the Higgs branch of the B-model.

It is worth noting that in fact we could get the matter content of the A-model without the 
use of an NS 5-brane in figure 1a.
This corresponds to the absence of the extra fundamental hypermultiplet (thin circle)
in the quiver diagram of the
B-model in figure 2b.
These theories are equivalent to the A and B models when the mass of the adjoint of the A-model
and the sum of the FI parameters of the B-model are zero, which is indeed the case here as
we will discuss later.
In the A-model, the relative position in the $(x^7,x^8,x^9)$ direction
of the 
D3 branes with respect to the NS 5-brane corresponds in the B-model to
the relative position in the $(x^3,x^4,x^5)$ direction of
the D3 branes with respect to the D5 brane. 
In other words, the parameters corresponding to the $U(1)$ part of the adjoint hypermultiplet
in the A-model correspond to  those of the vector multiplet of the diagonal $U(1)$ in the B-model.
This is the  D-brane picture of something discussed in the field theory context in \cite{dhoo},
 where the existence of a trivial
$R^4$ in the hypermultiplet moduli space of the A-model and in the vector multiplet
moduli space of the B-model has been pointed out.

Let us now recall some of the results of \cite{dhoo}.
Consider the A-model:
without mass terms, the vector multiplet moduli space is 
the $k$-symmetric product 
of an ALE space
\beq
{\cal M}_V({\rm A-model},\vm_{adj}=0, \vm_{fund}= 0)
  = Sym^k X_{A_{n-1}}
\stop
\label{symm}
\eeq
It has singularities inherited from 
the simple singularity of $A_{n-1}$ type of 
the ALE space $X_{A_{n-1}}$,
and also singularities coming from 
modding out by the action of the
symmetric 
group.
The masses for the fundamental hypermultiplets resolve 
the simple singularity of $X_{A_{n-1}}$.
The mass of the adjoint hypermultiplets resolves 
the quotient singularities of the symmetric product.
The other  effect of the mass terms is to lift some of the flat 
directions of the hypermultiplet moduli space. 
 
In the B-model, the resolution of the singularities of the
hypermultiplet moduli space
and  the lifting  of some of the flat directions for the vector 
multiplets
are caused by turning on FI terms. The  way in which the moduli 
spaces
are resolved or lifted  matches exactly with the 
A-model
when the vector multiplet and hypermultiplet moduli spaces
 are exchanged,
provided that the FI parameters are related to the mass 
parameters of the A-model in a certain way.

The mirror map between the mass parameters
of the A-model and the FI parameters of the B-model 
takes the form \cite{dhoo}
\beq
\vec{m}_i = \sum_{l=0}^i \vec{\zeta}_l,~~~~~~~~~~
\vec{m}_{adj} = \sum_{l=0}^{n-1} \vec{\zeta}_l
\comma
\label{map}
\eeq
where $\vec{m}_i$ are the masses of the fundamental 
hypermultiplets,
$\vec{m}_{adj}$ is the mass of the adjoint 
hypermultiplet and
$\vec{\zeta}_l$ are the FI parameters.
The freedom to shift the origin on the Coulomb branch of the
A-model has been used to choose $\vec{m}_{n-1}=\vec{m}_{adj}$.

In the brane configuration of figure 1a, the mass of the adjoint hypermultiplet is zero since the 
length of the open string stretching between the Dirichlet 3-branes is zero.
This corresponds in the dual theory to the case where
the sum of the FI parameters is zero. It is not immediately clear how to turn on a mass
for the adjoint hypermultiplet, or how to get a non-zero sum of the FI parameters
in the intersecting brane configurations we consider. A possibility might be
to turn on suitable string background fields, and it would be interesting to
investigate this point further. 

Since the $SL(2,Z)$ transformation exchanges the NS and Dirichlet 5-branes, it also
exchanges  $\vec{\omega}_i$ and $\vec{m}_{i}$.
Using the relation between  the FI parameters and $\vec{\omega}_i$  in (\ref{zeta})
we get the mirror map

\beq
\vec{m}_{i} - \vec{m}_{i-1}=  \vec{\zeta}_{i},~~~~~~~~~~i=1,...,n-1.
\label{mzeta}
\eeq
Equation (\ref{mzeta}) is equivalent to the mirror map (\ref{map})
with $\vm_{adj}=\sum_{l=0}^{n-1} \vec{\zeta}_l=0$. 

As indicated in (\ref{symm}), when the mass of the adjoint hypermultiplet
vanishes, the vector multiplet moduli space
of the A-model with gauge group $U(k)$ is a $k$-symmetric product 
of the vector multiplet moduli space of an $U(1)$ theory.
This can be read from figure 1a as follows:
A vev for the diagonal components of the 
adjoint hypermultiplet can separate the $k$ Dirichlet 3-branes in $(x^7,x^8,x^9)$,
and each of them may be considered as independent.
This implies that the  vector multiplet moduli space which is parameterized by the vev's
of the bosons in the vector multiplet in  $(x^3,x^4,x^5)$
is a product of the vector multiplet moduli spaces of each brane  modded by their permutation.

This  scenario
is  expected from another string theory viewpoint.
We can use D-branes  to probe the space-time geometry and 
the background gauge fields of string theory, where
enhanced gauge symmetry in the space-time theory is reflected in the D-brane 
world volume theory by enhanced global symmetry.
Consider a type I string theory  on $R^7\times T^3$. 
Performing a T-duality transformation on the $T^3$ coordinates we get the type I' theory:
type IIA with eight orientifolds and sixteen Dirichlet 6-branes and their images. The
probes are $k$ Dirichlet 2-branes of  the type IIA string.
When the probes are near $n$ coinciding Dirichlet 6-branes, as in figure 3, the worldvolume
theory of the probes is that of the A-model:
The gauge group is $U(k)$ as a consequence of having $k$ probes, the adjoint hypermultiplet arises
from the $D2-D2$ sector and the $n$ hypermultiplets in the fundamental arise from the
$D2-D6$ sector. Their masslessness reflects the $U(n)$ space-time gauge symmetry as
a global symmetry on the worldvolume of the probes.

\begin{figure}
\begin{center}
$\mbox{\epsfig{figure=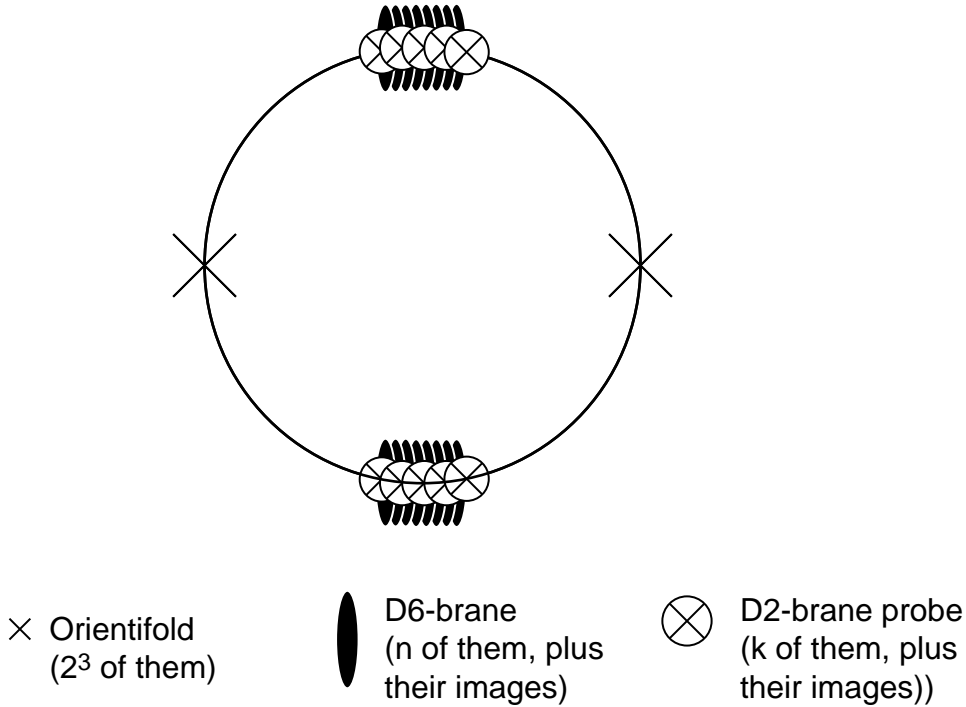}}$
\end{center}
\caption{$k$ D2-brane probes near $n$ D6-branes in type I'}
\end{figure}

A vev for the adjoint hypermultiplet separates the Dirichlet 2-branes. 
The vector multiplet moduli space of each brane can be determined  using the duality between 
M-theory on $R^7\times K_3$ and Type I string on $R^7\times T^3$ \cite{s}.
Under this duality each one of the Dirichlet 2-brane probes is mapped to an  M-theory
2-brane whose world volume is $R^3\times \{pt \in K_3\}$, which implies that its
vector multiplet moduli space is $K_3$.  More precisely, the vector multiplet moduli space
is determined by a neighborhood of the singularity in the $K3$ and  
is an ALE space of $A_{n-1}$ type. The symmetric product structure of the 
vector multiplet moduli space is a consequence of having $k$ separated Dirichlet 2-branes.
Analogous discussion has been presented for $Sp(k)$ gauge group in \cite{dhoo}
by taking the D2 and D6 branes to lie in an orientifold point,
and for $N=2$ gauge theories in four dimensions in \cite{asyt,dls}.

Consider now the role of instanton corrections.
When the adjoint hypermultiplet is massless we do not expect instanton corrections to the
metric on the vector multiplet moduli space, while we do expect them  when the 
adjoint is massive or absent \cite{dhoo}.
In the following we will verify this expectation
using the configurations of intersecting Dirichlet branes.
This will also enable us to get an understanding of the stringy origin of the two
types of coupling constants:
electric and magnetic in the terminology of \cite{hw}.

We expect the instanton corrections to  the vector multiplet
moduli space of the Dirichlet 3-brane
worldvolume theory to arise from monopoles which are instantons in three
dimensions. The monopoles in the D-brane picture arise from open D-string instantons
stretching between the Dirichlet 3-branes \cite{douglas}.  Open string
instantons are  holomorphic maps from the disc
to space-time such that the boundary of the disc is mapped to the Dirichlet brane
\cite{ooy,syz}.
Let us begin with a qualitative analysis.
In order to have open string instantons contributions in the A-model we need holomorphic maps 
from the disc to the cylindrical shaded area in figure 4a.

\begin{figure}
\begin{center}
$\mbox{\epsfig{figure=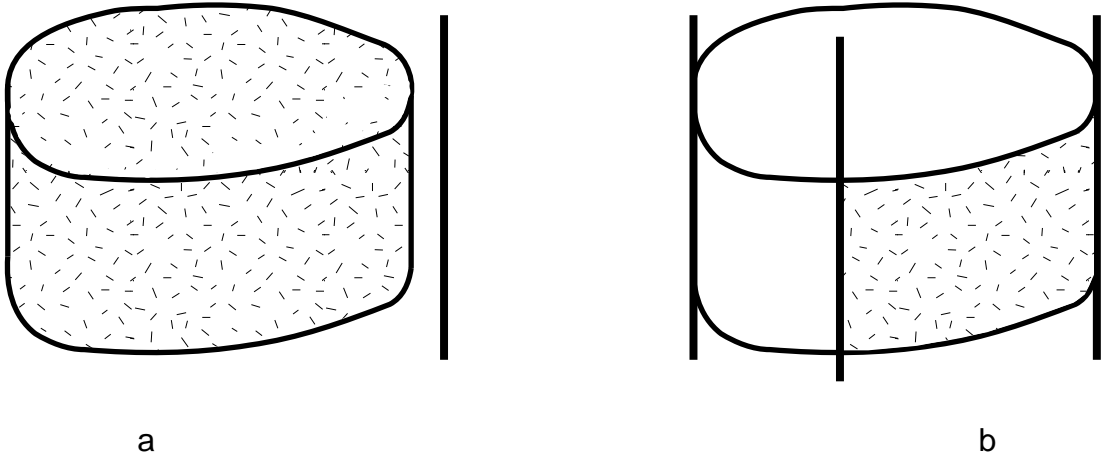}}$
\end{center}
\caption{The image in space-time of open string instantons.
There are no instanton corrections from figure 4a, but there are from figure 4b.}
\end{figure}

If we move the NS 5-brane as in figure 4a there is no such map. The reason is simple: if we 
think about the disc as a square then two of the edges are mapped
to two Dirichlet 3-branes. The other two edges have no boundary to be mapped to.
This is consistent with what we expect from field theory considerations \cite{dhoo}.
There is a delicate point in this discussion, since if we do not move the NS 5-brane,
the remaining edges can end in it.
However, we  do not expect to have instanton corrections in this case, as  will be discussed
in section 5.
We therefore conclude  that although the relevant holomorphic maps exist,
the coefficient of the contribution is zero.

Consider now for comparison the D-brane configuration in figure 5a.
\begin{figure}
\begin{center}
$\mbox{\epsfig{figure=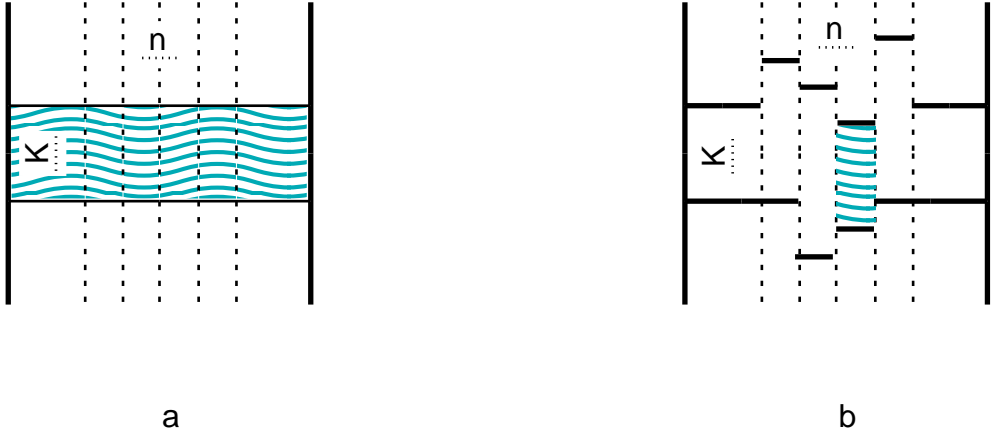}}$
\end{center}
\caption{An $U(k)$ gauge theory with $n$ fundamentals and no adjoint hypermultiplet is plotted
in its Coulomb branch
on the left and in its Higgs branch on the right.}
\end{figure}
This configuration yields on the Dirichlet 3-branes
a $U(k)$ gauge theory with $n$ hypermultiplets in the fundamental representation but without an adjoint.
In this case a typical open string instanton is a holomorphic map from the disc to the shaded
area in figure 4b, which has the topology of a disc. Such holomorphic maps exist,
suggesting that we will, as expected, get instanton corrections.
In section 5 we will confirm these results  from a different viewpoint.

Consider now the hypermultiplet moduli space corresponding to figure 5a which is depicted
in figure 5b. As we noted previously, it is determined
classically and in particular  is not corrected by instantons.
Naively, this seems to be in contradiction with the fact that there are holomorphic maps from the 
disc to the shaded area in figure 5b which has the topology of a disc.  
Recall however  that  
the instanton corrections are exponentials of the instanton action $\frac{|\vec{\phi}|}{g_m^2}$
where $\vec{\phi}$ corresponds to
the relative
position of the  Dirichlet 3-brane in  $(x^7,x^8,x^9)$ and  $\frac{1}{g_m^2}$ is the distance 
between the Dirichlet 5-branes \cite{hw}.
The   hypermultiplet moduli space is obtained in the limit
$\frac{1}{g_m^2}\rightarrow 0$  which means that the shaded area in figure 5b shrinks to  zero size,
and therefore
the instanton corrections are field independent, and should be taken into account at the classical
level.

Let us now discuss the relation between open string instantons and the Dirichlet 3-brane
worldvolume instantons in a more quantitative way.
First recall that the four dimensional gauge coupling is related to the string coupling
as
\beq
\frac{1}{g_4^2} = \frac{1}{g_{st}}
\comma
\label{gs}
\eeq
since the coefficient of the gauge kinetic 
term $F^2$ in the open string effective action is $\frac{1}{g_{st}}$.
This together with (\ref{R}) implies that
 \beq
\frac{1}{g_e^2} = \frac{r}{g_{st}}
\comma
\label{ge}
\eeq
where by $g_e$ we denote the dimensionful three dimensional gauge coupling.
We expect the "magnetic" coupling to be related to the electric one (\ref{ge})
by  an $SL(2,Z)$ transformation, leading to 
 \beq
\frac{1}{g_m^2} = r g_{st}
\stop
\label{gm}
\eeq
In the field theory limit $r\rightarrow 0, g_{st} \rightarrow 0$ the magnetic
coupling $\frac{1}{g_m^2}$ vanishes while the electric coupling $\frac{1}{g_e^2}$ can still be finite.
This is the traditional field theory corner of the moduli space of D-branes.

However, the above rough analysis clearly suggests that there are other parts of the moduli
space of D-branes.
Consider now the open D-string instanton corrections 
 to the vector multiplet moduli space of the 3-brane worldvolume
theory. These should take the form
\beq
\mbox{{\rm Open D-string Instanton }} \sim \exp \left(-\frac{A}{\alpha' g_{st}}\right)
\label{D}
\comma
\eeq
where $A$ is the area of the image of the instanton in space time.
The area is $r$ times the separation between the Dirichlet 3-branes which we denote
by $d$.
From field theory viewpoint the instanton contribution takes the form
\beq
\mbox{{\rm Field  Theory Instanton }}
 \sim \exp \left(-\frac{\langle \phi_V \rangle}{ g_e^2} \right)
\label{FD}
\comma
\eeq
where $\langle \phi_V \rangle$ is the vev for the scalars in the vector multiplet.  
This vev gives the masses 
of the W bosons through the Higgs mechanism.  
Since the  masses of the W bosons, which are associated with
the open strings stretching between the Dirichlet 3-branes, are given
by the length of the open strings times their tension we have
\beq
\langle \phi_V \rangle = \frac{d}{\alpha'}
\label{vev}
\stop
\eeq
Using (\ref{vev}) we see that indeed
(\ref{D}) and (\ref{FD}) are the same.

Consider now the open string instantons.
Their correction should take the form
\beq
\mbox{{\rm Open String Instanton }} \sim \exp \left( -\frac{A}{\alpha'}\right)
\label{Ds}
\stop
\eeq
From a "field theory" viewpoint the instanton contribution takes the form
\beq
\mbox{{\rm "Field Theory" Instanton }} \sim \exp \left(-\frac{\langle \phi_H\rangle}{g_m^2}\right)
\label{FDm}
\comma
\eeq
where $\langle \phi_H \rangle$ is related to the  vev's for the scalars in the hypermultiplet.
Using a reasoning similar to the one above, with open strings replaced by  D-strings, we have
\beq
\langle \phi_H \rangle = \frac{d}{\alpha' g_{st}}
\stop
\label{H}
\eeq
Using (\ref{H}) we see that  (\ref{Ds}) and (\ref{FDm}) are equal.

The above identification of the field theory instantons with the open string instantons provides one of the 
main direct links between the brane and the field theory moduli spaces. 
Indeed, the analysis yields the following picture: in the D-brane
moduli space there exists a part where the physics is described by a traditional
 field theory. In this case the
vector multiplet moduli space  can be corrected by instantons while the hypermultiplet
moduli space is determined classically. In addition,
there is a dual part in the moduli space, the "magnetic phase", where the hypermultiplet moduli
space is corrected, but the vector multiplet is not.
And finally there are regions were both moduli spaces 
get corrections and both the electric and the magnetic
couplings are finite.
The traditional field theory mirror symmetry does not explain 
what is the mirror
of a theory that has only a Coulomb branch but not a Higgs branch, since the mirror
theory has a Higgs branch but not a Coulomb branch \footnote{
This reminds of an analogous phenomena in the Calabi-Yau moduli space where the mirror
of a rigid Calabi-Yau manifold ($h_{21}=0$)  should have $h_{11}=0$ and is therefore not a 
Calabi-Yau manifold.}.
In order to describe such mirrors we have to enlarge the field theory
moduli space to the D-brane moduli space.

\subsection{Duality for $U(k)^n$ Gauge Groups}

Mirror symmetry for $U(k)^n$ gauge groups can be obtained from the
intersecting  D-brane configuration in figure 6a.
\begin{figure}
\begin{center}
$\mbox{\epsfig{figure=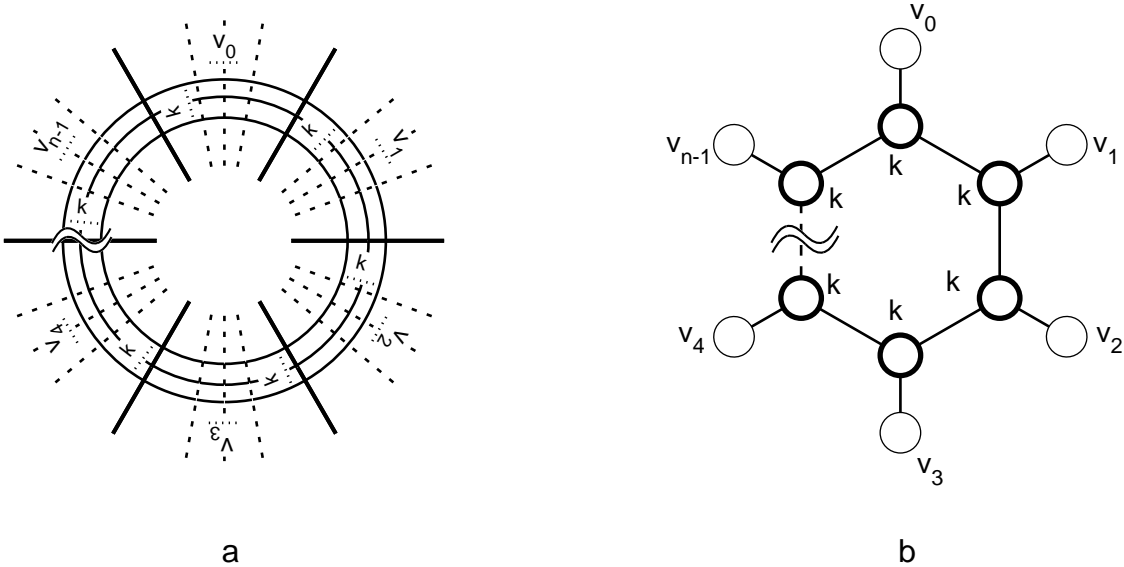}}$
\end{center}
\caption{The D-brane configuration of the A-model and the corresponding quiver diagram.}
\end{figure}
It consists of  $n$ NS 5-branes, $n$ sets of Dirichlet 5-branes,
each set containing $v_i,i=0,\ldots,n-1$ different Dirichlet 5-branes,
 and $k$ Dirichlet 3-branes.
Applying the same rules as in the previous section we see that
the three dimensional Dirichlet 3-brane worldvolume 
theory is an $N=4$ supersymmetric theory with gauge group
and matter content associated with the quiver diagram 6b.
This will be called the A-model.
The  gauge
group is $U(k)^n$, one $U(k)$ for each node of the extended Dynkin diagram.
There  are two kinds of matter. As before,
for each pair of gauge groups whose nodes are
connected by an edge there is matter transforming as $({\bf k},{\bf k}^{\ast})$ under 
$U(k)\times U(k)$. In addition, there are $v_i$ matter fields transforming
in the fundamental representation of the $i$th $U(k)$ gauge group.

We denote the A-model as $(U(k)^n;\{v_i\})$. 
The Higgs branch of the A-model is  the moduli space of instantons
 on a vector bundle $V$ over an ALE space of type $A_{n-1}$. More precisely,
it describes the moduli space ${\cal M}_k(V)$
of instantons of instanton number $k$ on
 $V=\oplus {\cal R}_i^{\otimes v_i}$, with gauge group $U(V)$,
where ${\cal R}_i$ are particular line
bundles 
over the ALE space associated to the different representations of ${\bf Z}_n$
\cite{kn}.

Performing an $SL(2,Z)$ transformation on the D-brane configuration in figure 6a
yields  a configuration 
which corresponds to the quiver diagram of figure 7 in the case where
all $v_i \geq 1$.

\begin{figure}
\begin{center}
$\mbox{\epsfig{figure=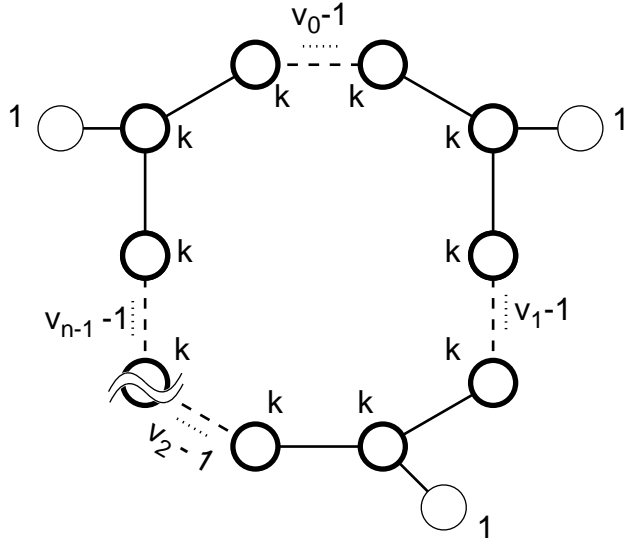}}$
\end{center}
\caption{The quiver diagram of the B-model.}
\end{figure}

In the notations of \cite{dhoo}, the
 B-model gauge theory in figure 7 is $(U(k)^m;\{w_i\})$, where
\beq
\sum_{i=0}^{n-1} v_i = m,~~~~~~~\sum_{i=0}^{m-1} w_i = n
\stop
\eeq

The mirror map takes the following form: Denote by $\vm_i^{(B)}$, $\sum_{l=0}^{j-1} w_l \leq i 
< \sum_{l=0}^j w_l$  the masses
of the hypermultiplets in the B-model charged only under the $j^{\rm th}$
$U(k)$. 
In addition, there are $m$ masses of hypermultiplets charged under
two $U(k)$'s. Using the freedom to shift the origin on the Coulomb branch,
we can choose all these masses equal to the same value which we denote by
$\vm_{2f}^{(B)}$, and in addition we can choose $\vm_{n-1}^{(B)}=0$. 
Then the relation between the FI parameters $\vec{\zeta}_i^{(A)}$ of the
A-model and the masses of the B-model reads \cite{dhoo}
\bea
\sum_{l=0}^{i} \vec{\zeta}_l^{(A)} & = & \vm_i^{(B)} + (\sum_{l=0}^{i} v_l) \vm_{2f}^{(B)} 
\nonumber \\{}
\sum_{l=0}^{n-1} \vec{\zeta}_l^{(A)} & = & (\sum_{l=0}^{n-1} v_l) \vm_{2f}^{(B)}.
\label{mrr}
\eea
In our case $\vm_{2f}^{(B)}$ vanishes since the open string stretching between Dirichlet 3-branes
which gives rise to
a hypermultiplet charged under
two $U(k)$'s is of zero length at the origin of the Coulomb branch.
This is the analog of the adjoint hypermultiplet of the previous sections.
Similarly, the  sum of the FI parameters in the A-model is zero, which can be
easily seen from (\ref{zeta}).

As before, the $SL(2,Z)$ transformation replaces the NS and Dirichlet 5-branes, and exchanges 
$\vec{\omega}_i$ and $\vec{m}_{i}$.
Using (\ref{zeta}), the FI parameters of the A-model are related to the mass parameters of the B-model
by
\beq
\vec{m}_{i}^{(B)} - \vec{m}_{i-1}^{(B)}=  \vec{\zeta}_{i}^{(A)},~~~~~~~~~~i=1,...,n-1.
\label{mz1}
\eeq
Equation (\ref{mz1}) is equivalent to the mirror map (\ref{mrr})
with $\vm_{2f}^{(B)}=0$.

\section{Duality for $\prod_iU(k_i)$ Gauge Groups}

In this section we generalize the examples of the previous section to dual models based on
 $\prod_iU(k_i)$ gauge groups.
Consider the quiver diagram in figure 8a, which encodes the field content
of a $\prod_iU(k_i)$ gauge theory.

\begin{figure}
\begin{center}
$\mbox{\epsfig{figure=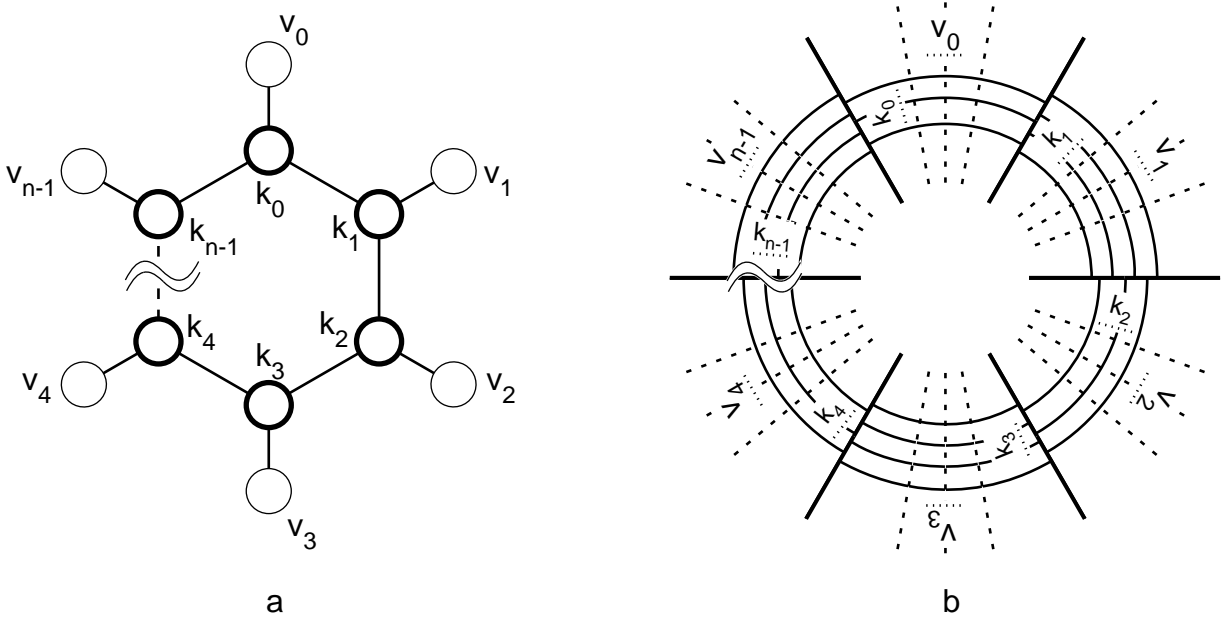}}$
\end{center}
\caption{The D-branes configuration of the A-model and the corresponding quiver diagram.}
\end{figure}

We will analyze the mixed branches of this theory and find the criterion for complete Higgsing,
both
from a field theory viewpoint and via a consideration of D-brane configurations. 
This will enable us to find  the procedure to construct the mirror partner of the model. 

Denote  the internal and external indices of the quiver diagram by the vectors
$\bk=(k_0,\ldots,k_{n-1})$ and $\bv=(v_0,\ldots,v_{n-1})$  respectively. 
For a given choice of FI parameter 
$\zeta=(\vec{\zeta}_0,\ldots,\vec{\zeta}_{n-1})$
the hypermultiplet moduli space of the gauge theory is  
the quiver variety $\M_{\zeta}(\bk,\bv)$ which is a certain moduli space of instantons. 
When 
$\zeta$ is chosen to be generic, the gauge group is completely 
broken and
$\M_{\zeta}(\bk,\bv)$ is smooth. In this case, the theory only has a Higgs phase.

\subsection{Mixed Branches in the Model with $\zeta=0$, $m=0$}

When the FI parameters $\zeta$ and the bare masses $m$
are not generic, the moduli space of vacua consists
of several branches touching each other at phase 
transition 
points (or lines, surfaces etc).
We consider here the most special case $\zeta=0$, $m=0$.
A mixed branch is characterized by a (conjugacy class 
of the)
unbroken gauge group $G$. The classification of the possible 
unbroken gauge groups $G\subset \prod_i U(k_i)$ 
is
given in section 6 of \cite{Na1} and section 3 and 4 of 
\cite{Na2}.
According to these, $G$ always takes the following form:
\beq
G=\prod_{a=1}^{p}U(\kappa_a)\times\prod_{i=0}^{n-1} 
U(\ell_i)
\stop
\label{unbroken}
\eeq
Here $U(\kappa_a)$ is the diagonal subgroup of 
$\prod_{i=0}^{n-1} U(\kappa_a)_i \subset \prod_{i=0}^{n-1} U(k_i)_i$,
and $U(\ell_i)$ is a subgroup
of $U(k_i)_i$.
Let us denote $\tilde{k}_i=k_i-\sum_a\kappa_a-\ell_i$.
Then, the mixed branch with unbroken gauge group (\ref{unbroken})
is the product
${\cal H}_{({\bf \tilde{k}},\kappa)}
\times {\cal V}_{({\bf \tilde{k}},\kappa)}$,
where ${\cal V}_{({\bf \tilde{k}},\kappa)}$ is a space 
of (quaternionic) dimension
$\sum_a\kappa_a+\sum_i\ell_i$ and ${\cal H}_{({\bf 
\tilde{k}},\kappa)}$
is given by
\beq
\M_0^{\rm reg}({\bf \tilde{k}},\bv)\times
Sym^{\kappa} (\C^2/\Z_n-\{0\})
\comma
\eeq
where
$\M_0^{\rm reg}({\bf \tilde{k}},\bv)$ is the completely 
Higgsed phase
of the model with internal indices ${\bf 
\tilde{k}}=(\tilde{k}_0,
\ldots,\tilde{k}_r)$ and external indices $\bv$. 
$Sym^{\kappa}X$ is the subspace of the symmetric 
product
$Sym^{|\kappa|}X$, $|\kappa|=\sum_a\kappa_a$, consisting 
of configurations
of ${|\kappa|}$-points in $X$ where $\kappa_a$ of them
are in the same position.

We can find a corresponding branch in a three and 
5-brane configuration.
We assume that the nonemptyness of $\M_0^{\rm reg}({\bf k},\bv)$ 
corresponds to the
existence of a complete Higgs phase in the brane picture.
It will be shown in the next subsection that this is indeed correct.
The branch consists of configurations of Dirichlet 3-branes
where $\ell_i$ of the $k_i$ Dirichlet 3-branes in the $i^{\rm 
th}$ interval
between the NS 5-branes are constrained to end on the NS 5-branes,
 and there are 
$|\kappa|$ Dirichlet 3-branes
without boundary moving freely from the NS 5-branes but 
$\kappa_a$ of them
($a=1,\ldots,p$)
are constrained to have the same position in the 
($x^7,x^8,x^9$) direction. In addition, there are other Dirichlet 3-branes, that
are partly generated
by passing Dirichlet 5-branes through NS 5-branes, as will be explained in detail
in the following.

\subsection{ Criterion for Existence of Complete Higgs Phase}

In \cite{dhoo} we proposed a 
condition
for the existence of a complete Higgs phase in the models corresponding to
quiver diagrams with indices $(\bk,\bv)$.
It is given by (see e.g. equation (8.5)) 
\beq
v_i\geq 2k_i-k_{i-1}-k_{i+1}.
\label{convexity}
\eeq
In fact  the same issue is discussed in \cite{Na1} and solved 
in  \cite{Na2}.
The criterion (Corollary 10.9 of \cite{Na2}) is indeed (\ref{convexity}) provided that the condition
\beq
\sum_{i=0}^{n-1}v_i\geq 2
\label{cond2}
\eeq
holds. 
We now derive (\ref{convexity}) and 
(\ref{cond2})  from the three and 
5-brane picture.
In order to get a 
complete Higgs phase we have to  take all the 
3-branes off the
NS 5-branes and constrain  them to end on the Dirichlet 
5-branes.
We do this as in \cite{hw} by moving Dirichlet 
5-branes through
NS 5-branes creating new Dirichlet 3-branes between 
them.
Let us focus on the neighborhood of the $i^{\rm th}$ 
interval of NS
5-branes in figure 9. 
\begin{figure}
\begin{center}
$\mbox{\epsfig{figure=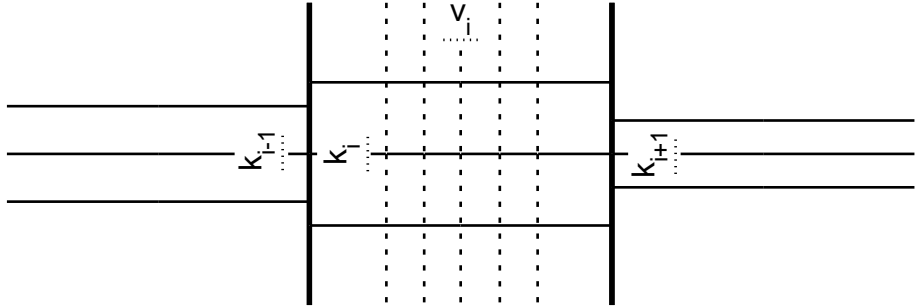}}$
\end{center}
\caption{The $i^{\rm th}$ 
interval of NS 5-branes.}
\end{figure}

We distinguish the following three 
cases:
(i) $k_i\geq k_{i-1}, k_{i+1}$,
(ii) $k_{i+1}\geq k_i\geq k_{i-1}$ and
(iii) $k_i\leq k_{i+1},k_{i-1}$.
We will derive the condition for taking the Dirichlet 3-branes off 
the NS 5-branes in the order
(i)$\Rightarrow$(ii)$\Rightarrow$(iii)
inductively.
In case (i) we move $k_i-k_{i-1}$ Dirichlet 
5-branes in the interval
to the left-next interval and $k_i-k_{i+1}$ to the 
right-next interval.
This is possible if and only if $v_i\geq 
(k_i-k_{i-1})+(k_i-k_{i+1})$
which is the condition (\ref{convexity}).
In  case (ii), given that the Dirichlet 5-branes have been moved as in step (i), 
$k_{i+1}-k_i$ Dirichlet 5-branes 
enter in the interval
from the right-next one. So now, there are 
$v_i+k_{i+1}-k_i$ Dirichlet 5-branes.
Then, we only have to move $k_i-k_{i-1}$ of them to the 
left-next interval,
which is possible if and only if
$v_i+k_{i+1}-k_i\geq k_i-k_{i-1}$.
Finally in the case (iii) where the condition 
(\ref{convexity}) is trivially
satisfied, given (i) and (ii), $k_{i+1}-k_i$
Dirichlet 5-branes enter from the right while $k_{i-1}-k_i$ 
from the left.
At this stage it is obvious that all the Dirichlet 3-branes can 
be taken off the
NS 5-branes. 
Now we have to find a phase where all the Dirichlet 3-branes 
are constrained 
to end on Dirichlet 5-branes. If there is no Dirichlet 
5-brane,
it is impossible. If there is only one Dirichlet 
5-brane as in figure 10,
even though all the 3-branes can intersect with it,
we can easily take them
off because none of them has a boundary, and there is
no complete Higgs phase.
If there are two or more Dirichlet 5-branes as in figure 11, then,
there is obviously a phase where all the 3-branes end 
on Dirichlet
5-branes at their boundaries which cannot be taken off 
the Dirichlet 5-branes. 
Thus, we rederived the condition
(\ref{cond2}).
\begin{figure}
\begin{center}
$\mbox{\epsfig{figure=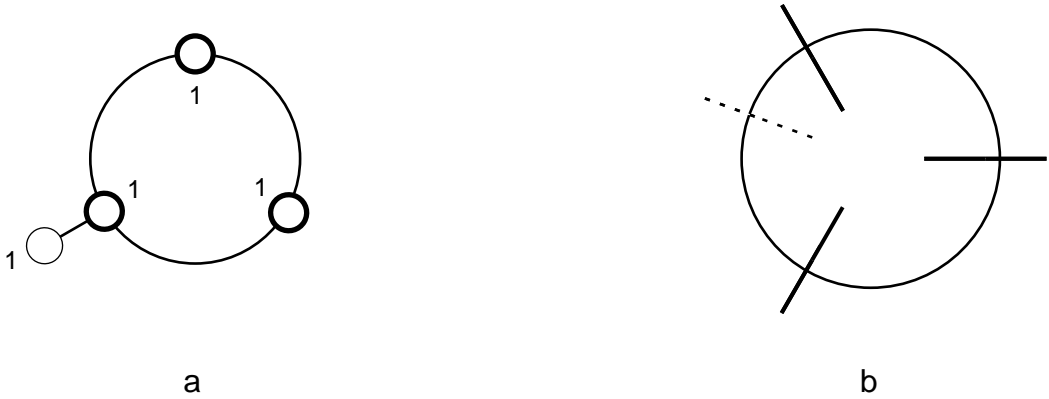}}$
\end{center}
\caption{A configuration with one Dirichlet 5-brane where there is no complete Higgsing
since the condition (\protect\ref{cond2}) is not satisfied.}
\end{figure}

\begin{figure}
\begin{center}
$\mbox{\epsfig{figure=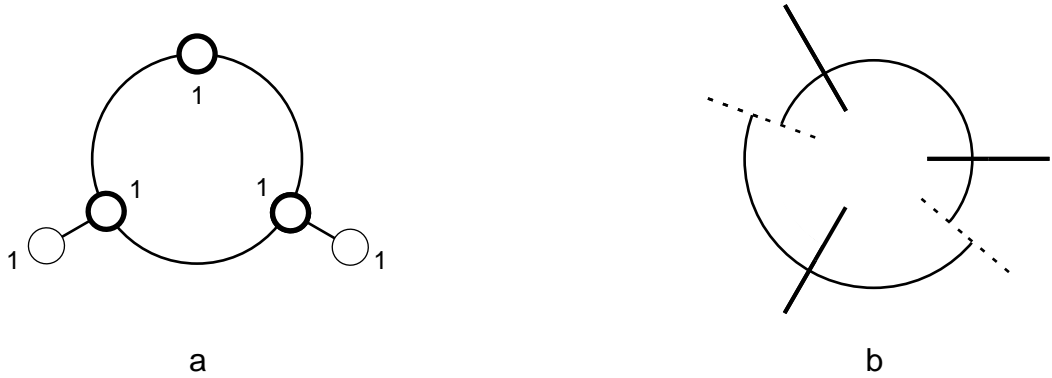}}$
\end{center}
\caption{A modification of figure 10,  with two Dirichlet 5-branes where there is a
 complete Higgsing
since both conditions (\protect \ref{convexity}) and (\protect\ref{cond2}) are satisfied.}
\end{figure}

As an application of the condition for complete Higgsing, we will show that when
two Dirichlet
5-branes  pass each other the field theory may undergo a phase transition in the D-brane
moduli space. To be precise,  5-branes of the same type do not have to pass each other but can be exchanged
by moving them in the three coordinates transverse to their worldvolume.

Consider the quiver diagram of figure 12a.
The intersecting brane configuration corresponding to it can be brought by the usual
rule for Dirichlet 5-branes passing an NS 5-brane to the form of 
figure 12c.
If we allow a Dirichlet 5-brane to pass another 
Dirichlet 5-brane we get the brane configuration
of figure 13a.
\begin{figure}
\begin{center}
$\mbox{\epsfig{figure=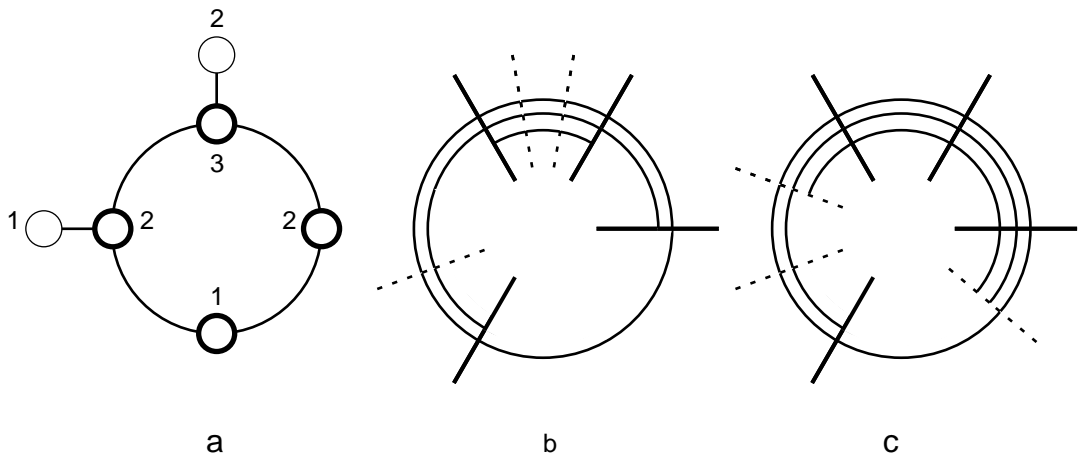}}$
\end{center}
\caption{A quiver, its Coulomb branch description via D-branes in 12b, and its
 Higgs branch description
in 12c.}
\end{figure}
\begin{figure}
\begin{center}
$\mbox{\epsfig{figure=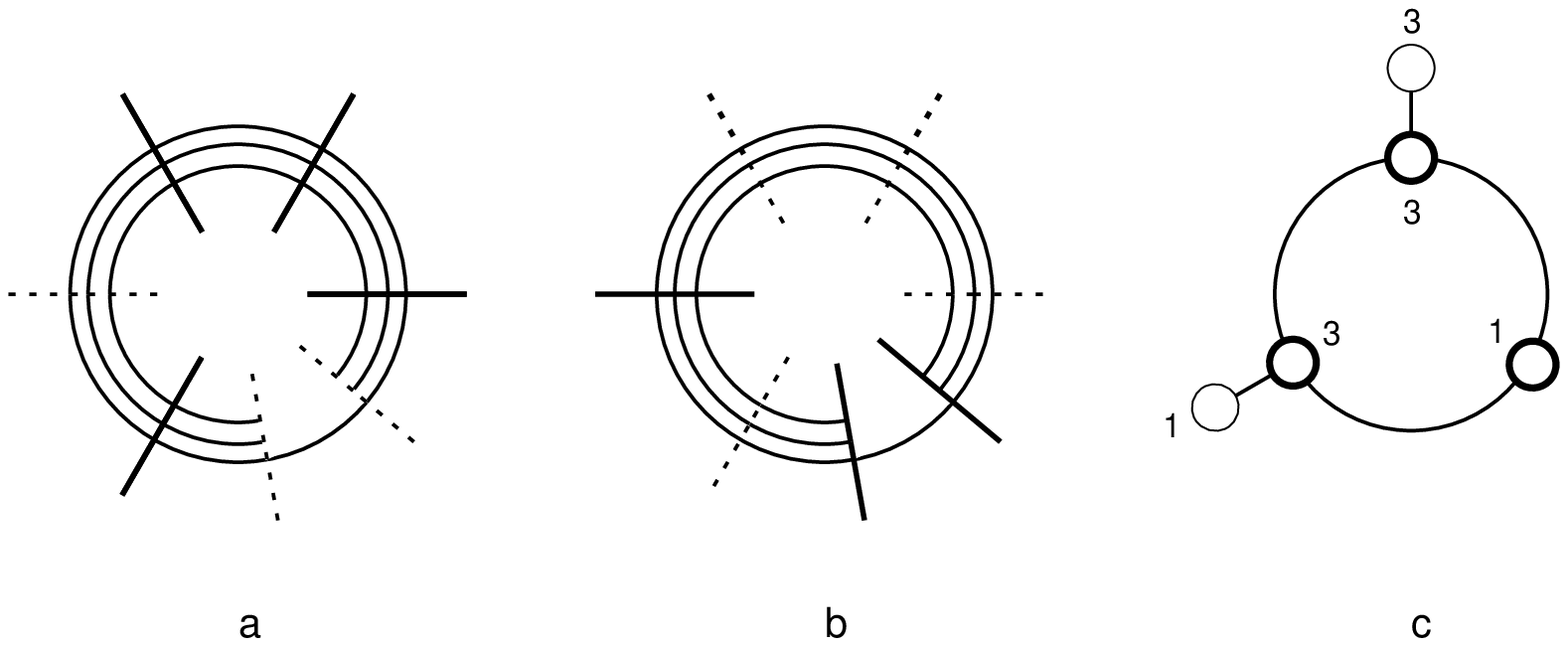}}$
\end{center}
\caption{The D-brane configuration 13a is derived from figure 12c if we we allow 
a Dirichlet 5-brane to pass another Dirichlet 5-brane. 13b is the $SL(2,Z)$ dual
of 13a. The quiver in 13c corresponding to 13b is a false mirror of 12a. }
\end{figure}

 Using $SL(2,Z)$ transformations on  figure 13a we will conclude that 
the quiver diagram of figure 13c is the mirror of the quiver diagram of figure 12a.
However, this is incorrect since the model of figure 12a has a Coulomb phase
while the model of figure 13c does not have a complete Higgs phase, since the condition
(\ref{convexity}) is not satisfied.
One can also  check this by  counting the dimensions of the vector multiplet
and hypermultiplet moduli spaces of the models.
Therefore we are
led to a contradiction. This implies that when  two 5-branes of the same 
type  pass each other, the field theory 
content on the D3 brane is changed --- we have a nontrivial 
phase transition on its worldvolume.  This  phenomenon 
has a simple interpretation in terms of D-branes.  
The coupling contants, as well as the Fayet-Iliopoulos 
and mass parameters of the D3 brane worldvolume theory are determined by 
the positions of the various 5-branes.  Varying them is simply 
moving around in the 5-brane moduli spaces.  Incidentally,  
these moduli spaces also have a gauge theoretic interpretation, 
i.e. moduli spaces of the supersymmetric gauge theories on the 
5-brane worldvolumes.
D-branes thus link different field theories as phases 
corresponding to different regions of the brane moduli 
space.

This is one of the most important concepts emerging from 
this  construction.
It  has much  in common with  the conifold 
transition \cite{strom}. Here we encountered phase transitions 
in field theory while moving smoothly
in the D-brane moduli space. The conifold transition 
is a phase transition from the supergravity
field theory viewpoint and is smooth  
in the  closed string theory.
 
\subsection{Mirror Pairs}
The above consideration of three and 5-brane 
configurations
leads to a recipe for the construction of mirror pairs.
The mirror of the model based on affine $A_{n-1}$ 
Dynkin diagram with indices $(\bk,\bv)$ is the model
based on affine $A_{m-1}$ Dynkin diagram with indices 
$(\bl,\bw)$,
$\bl=(l_0,\ldots,l_{m-1})$, $\bw=(w_0,\ldots,w_{m-1})$,
which are given in the following way.
Without loss of generality, we may assume that
$k_0=\max\{k_i\}$.
Let us introduce the notations
${\bf \tilde{v}}=\bv-{\bf C}_n\bk$ and
${\bf \tilde{w}}=\bw-{\bf C}_m\bl$ where ${\bf C}_n$ and 
${\bf C}_m$
are the Cartan matrices for the affine Lie algebras
$\widehat{sl(n)}$ and $\widehat{sl(m)}$ respectively. 
Then, $\bw$ and ${\bf \tilde{w}}$ are given by
\beqa
\bw&=&(1,\underbrace{0,\ldots,0}_{\tilde{v}_1-1},1,
\underbrace{0,\ldots,0}_{\tilde{v}_2-1},1,
\ldots,1,
\underbrace{0,\ldots,0}_{\tilde{v}_{n-1}},1,
\underbrace{0,\ldots,0}_{\tilde{v}_0-1})\\
{\bf \tilde{w}}&=&(
\underbrace{0,\ldots,0}_{k_0-k_1},1,
\underbrace{0,\ldots,0}_{v_1-1},1,
\underbrace{0,\ldots,0}_{v_2-1},1,
\ldots,1,
\underbrace{0,\ldots,0}_{v_{n-1}},1,
\underbrace{0,\ldots,0}_{v_0+k_1-k_0-1}).
\eeqa
Here, if there is a space of negative length $v_i-1=-1$, 
the left-next entry is added to the right next entry.
If the negative space appears in the right extreme,
the left-next entry is added to the first
entry. Given $\bw$ and ${\bf \tilde{w}}$, the vector 
$\bl$ is determined up to
an ambiguity of adding $(1,1,\ldots,1)$. This is fixed 
by saying that
$l_0=k_0$. Note that $n$ and $m$ are related by
\beq
m=\sum_{i=0}^{n-1}v_i,
\qquad
n=\sum_{\imath=0}^{m-1}w_{\imath}
\eeq
We can associate to vectors of indices Young diagrams
$\bv, {\bf \tilde{v}}, \bw, {\bf \tilde{w}}$ $\to$
$Y(\bv), Y({\bf \tilde{v}}), Y(\bw), Y({\bf 
\tilde{w}})$.
Here, $Y(\bv), Y({\bf \tilde{v}})$ are Young diagrams whose
rows have lengths $\sum_{i=0}^j v_i, j=0,\ldots,n-1$ and
$\sum_{i=0}^j \tilde{v}_i, j=0,\ldots,n-1$. The Young diagrams
$Y(\bw), Y({\bf\tilde{w}})$ have columns of lengths 
$\sum_{i=q-j}^q w_i, j=0,\ldots,m-1$ and 
$\sum_{i=\tilde{q}-j}^{\tilde{q}} \tilde{w}_i, j=0,\ldots,m-1$,
where $q$ and $\tilde{q}$ are the largest indices such
that $w_q \neq 0$, $\tilde{w}_{\tilde{q}}\neq 0$.
The Young diagrams are related by
\beq
Y(\bv)={}^tY({\bf \tilde{w}}),
\qquad
Y(\bw)={}^tY({\bf \tilde{v}}),
\label{transposition}
\eeq
where ${}^tY$ is the transposition of $Y$ along the 
NorthWest-SouthEast
diagonal.
This generalizes the transposition rule \cite{dhoo}
of the models with common internal indices in which
${\bf \tilde{v}}=\bv$, ${\bf \tilde{w}}=\bw$.

As an illustration, consider gauge theory associated to
the quiver diagram given in
figure 14a. One can find its mirror by considering the 
corresponding brane
configuration, moving D5 branes in an appropriate 
way, and performing
an $SL(2,{\bf Z})$  transformation. The result is given 
in figure 14b, the corresponding Young diagrams in
figure 15 and 16. 
\begin{figure}
\begin{center}
$\mbox{\epsfig{figure=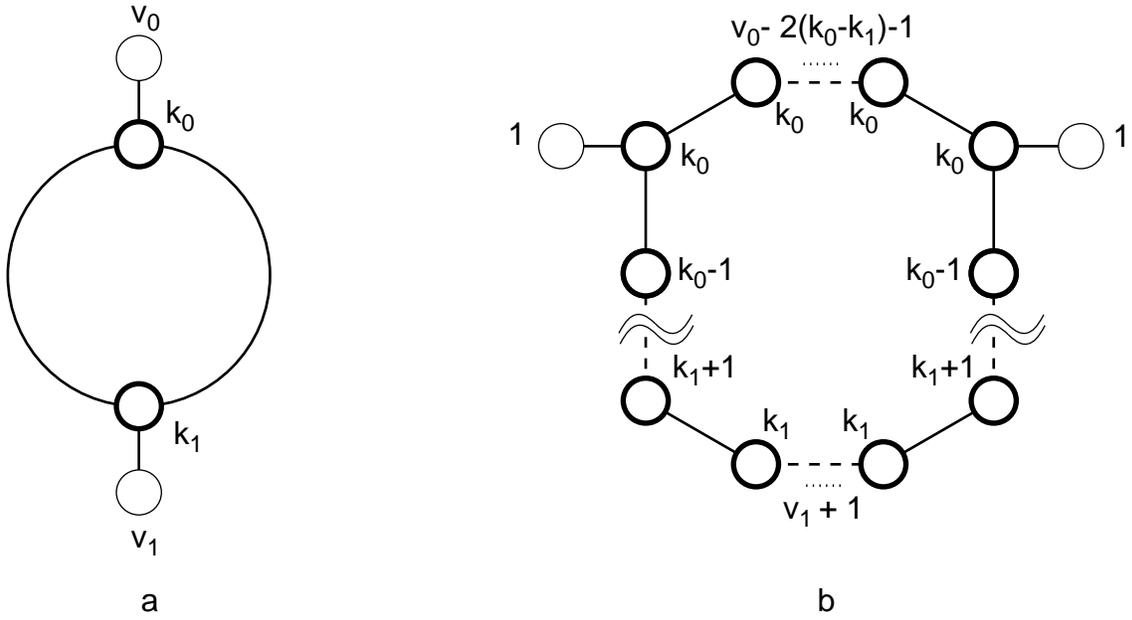}}$
\end{center}
\caption{An example of a mirror pair.}
\end{figure}

\begin{figure}
\begin{center}
$\mbox{\epsfig{figure=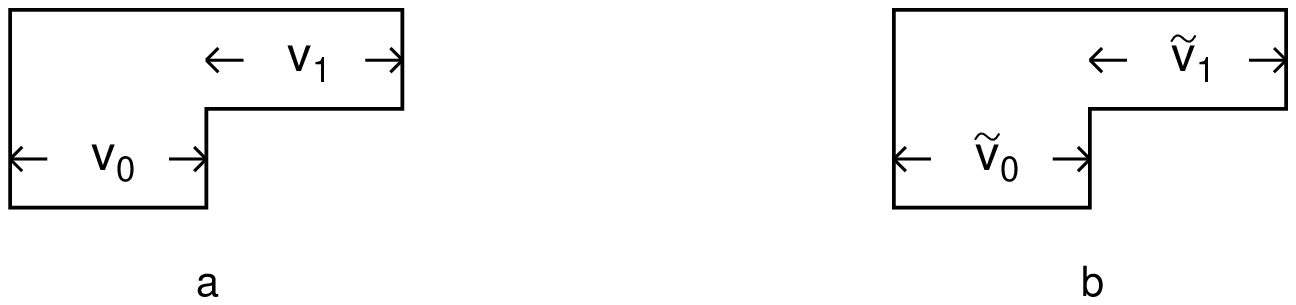}}$
\end{center}
\caption{Young diagram for the model in figure 14a: Figure 15a is $Y(\bv)$ and figure 15b is
$Y({\bf\tilde{v}})$
.}
\end{figure}

\begin{figure}
\begin{center}
$\mbox{\epsfig{figure=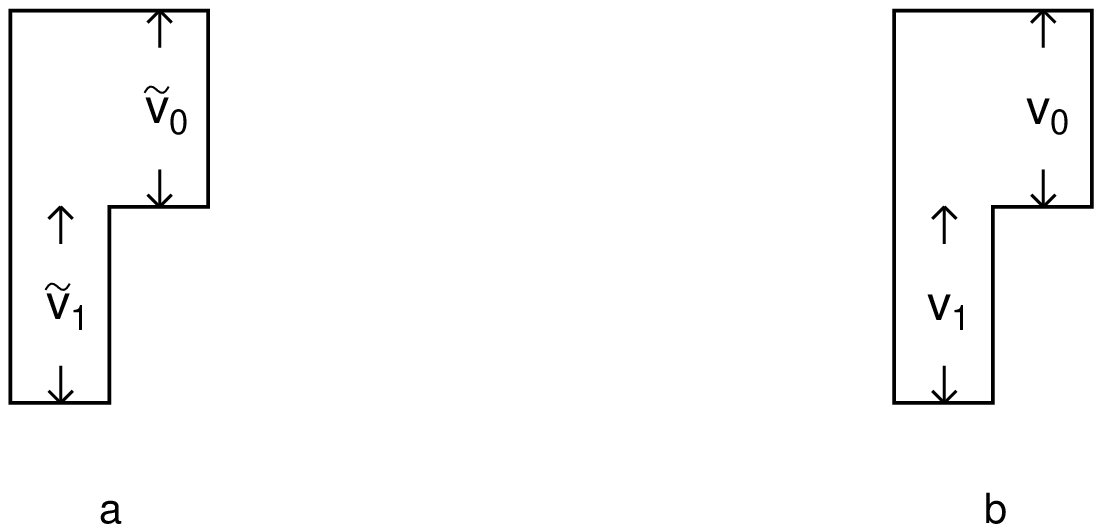}}$
\end{center}
\caption{Young diagram for the model in figure 14b: Figure 16a is $Y(\bw)$ and figure 16b is
$Y({\bf\tilde{w}})$
.}
\end{figure}

\subsection{Mirror Symmetry and Level-Rank Duality}

In \cite{Na1,Na2} Nakajima showed 
that there is an action of the affine Lie algebra 
$\widehat{sl(n)}$ on the middle dimensional
homology groups ${\rm H}_{mid}(\M_{\zeta}(\bk,\bv))$ of the 
moduli spaces $\M_{\zeta}(\bk,\bv)$
of the models based on the affine $A_{n-1}$ Dynkin diagram 
where $\zeta$ is chosen to be generic.
More precisely,
{\it for a fixed $\bv$, there is an action of the affine 
Lie algebra
$\widehat{sl(n)}$ on the vector space
$$\bigoplus_{\bk}{\rm H}_{mid}(\M_{\zeta}(\bk,\bv))$$
where $\bk$ runs over all possible internal indices.
It is the irreducible integrable highest weight 
representation
of highest weight $\Lambda_{\bv}$ where
${\rm H}_{mid}(\M_{\zeta}(\bk,\bv))$
is the weight space for the weight
$\Lambda_{\bv}-\alpha_{\bk}$} (Theorem 10.16 of 
\cite{Na1},
Theorem 10.3 of \cite{Na2}).
The symbols $\Lambda_{\bv}$ and $\alpha_{\bk}$ in the 
statement
are defined by
$\Lambda_{\bv}=\sum_{i=0}^rv_i\Lambda_i,$
$\alpha_{\bk}=\sum_{i=0}^r k_i\alpha_i,$,
where $\Lambda_i$ are the fundamental affine weights and 
$\alpha_i$
are the simple affine roots.
It is interesting to note that the middle dimensional 
homology of the Higgs branch of the model based on 
the affine
$A_{n-1}$ Dynkin diagram with indices $(\bk,\bv)$
is a weight space of a representation of 
$\widehat{sl(n)}$ at level $m$ while
the one of the mirror (based on the
affine $A_{m-1}$ Dynkin diagram with indices
$(\bl,\bw)$) is a weight space of a representation of 
$\widehat{sl(m)}$ at level $n$.
Indeed the condition (\ref{convexity}) for existence of 
complete Higgs phase
in the model with $\zeta=m=0$
simply says that $\Lambda_{\bv}-\alpha_{\bk}$ {\it is 
non-negative},
which implies that $\Lambda_{\bv}-\alpha_{\bk}$ is a 
weight of the integrable representation
of highest weight $\Lambda_{\bv}$ (see \cite{Kac} 
Proposition 12.5).
The other condition (\ref{cond2}) shows that the 
mirror is also based on
an affine Dynkin diagram. 
This strongly reminds us of the so called level-rank 
duality in two dimensional conformal field theories,
solvable statistical models, and quantum groups
(see \cite{NT} and references therein).
Usually, level-rank duality is stated in terms of 
transposition
of Young diagrams which can be contrasted with 
(\ref{transposition}).
It would be interesting to pursue the relation 
between mirror symmetry
and level-rank duality further and find its physical 
interpretation. In particular, we would like to
know the meaning of the curious fact that the mirror
symmetry does not map highest weights
to highest weights, and whether the middle-dimensional
homologies of the moduli spaces have any interpretation
in field theory. 

\section{Abelian Dual Pairs}

In this section we describe a large class of dual pairs of Abelian
$N=4$ gauge theories with matter. In fact, modulo some subtleties,
we will be able to find the dual of any given Abelian gauge theory
with matter, as long as there is some matter charged under each 
$U(1)$ gauge group. If this would not be the case, part of the
theory would be a pure $N=4$ $U(1)$ Yang-Mills theory. The moduli
space of this theory contains only a Coulomb branch with metric
(for notations see \cite{dhoo})
\be
ds^2 = \frac{1}{e^2} d\vec{r}^2 + e^2 d\phi^2
\ee
which describes a cylinder ${\bf R}^3 \times S^1_e$, where
$S^1_e$ is a circle with radius $e$, $e$ being the gauge coupling.
One could argue that in the ``infrared'' limit $e^2\rightarrow \infty$
this metric becomes the flat metric on ${\bf R}^4$, as the circle
get decompactified. The dual theory of pure $U(1)$ Yang-Mills theory
is then easily found, it is a theory with one free hypermultiplet and
no gauge group, whose moduli space contains just a Higgs branch
${\bf R}^4$. We will in what follows never consider this trivial
mirror pair
\be
\mbox{pure $U(1)$ gauge theory} \leftrightarrow \mbox{neutral hypermultiplet},
\ee
but one can always add arbitrary many copies of it to the Abelian
dual pairs that we describe below.

Let us now take any Abelian gauge theory with gauge group $U(1)^{N_c}$ and
with $N_f$ flavors, and assume that it does not contain pure gauge groups
or neutral hypermultiplets. We further assume that after a change of basis
the hypermultiplets can be arranged in two sets $Q_i$, $i=1\ldots N_c$ and
$Q_{\alpha}$, $\alpha=1 \ldots N_f-N_c$, such that $Q_i$ has charge $1$ under
the $i$th  $U(1)$ and is neutral with respect to the others. The
charge of $Q_{\alpha}$ under the $i$th $U(1)$ gauge group can be
an arbitrary integer which we denote by $m_{i\alpha}$. This situation 
cannot always be achieved, take for example one $U(1)$ with two
hypermultiplets with charges $2$ and $3$. We will discuss the general 
situation later. 

The total charge matrix for the hypermultiplets $Q_i$ and $Q_{\alpha}$ 
is the $N_c \times N_f$ matrix $({\bf 1} | m)$. We now claim that the 
dual of this theory is the $N=4$ gauge theory with gauge group
$U(1)^{N_f-N_c}$ and $(N_f-N_c) \times N_f$ charge matrix $({\bf 1}| -m^t)$,
\be
U(1)^{N_c},\mbox{\rm charges } ({\bf 1}|m) 
\leftrightarrow
U(1)^{N_f-N_c},\mbox{\rm charges } ({\bf 1}|-m^t).
\ee
This is slightly reminiscent of the non-Abelian duality in
$N=1$ theories in four dimensions \cite{sei}, where the dual
theories have gauge groups $SU(N_c)$ and $SU(N_f-N_c)$. 

As a first check, we compute the dimensions for the Coulomb and
Higgs branches. The first model has a Coulomb branch of (quaternionic)
dimension $N_c$, and a Higgs branch of dimension $N_f-N_c$, while
the second model has these numbers interchanged, as it should. Furthermore,
the first model has $N_c$ Fayet-Iliopoulos parameters $\vec{\zeta}^A_i$, and
$N_f$ mass parameters. We can use the freedom to choose the origin in
the Coulomb branch to choose the mass parameters of $Q_i$ to be zero,
so that all what remains is the $N_f-N_c$ mass parameters $\vec{m}^A_{\alpha}$ of
the $Q_{\alpha}$. The dual theory has $N_f-N_c$ Fayet-Iliopoulos parameters
$\vec{\zeta}^B_{\alpha}$ and $N_c$ independent mass parameters
$\vec{m}^B_i$. The number of Fayet-Iliopoulos and mass
parameters is indeed interchanged under the duality. In fact, we
will demonstrate later that the precise mirror map is simply
\be \label{mirrormap}
\vec{\zeta}^A_i \leftrightarrow \vec{m}^B_i, \qquad
\vec{m}^B_{\alpha} \leftrightarrow \vec{\zeta}^A_{\alpha}.
\ee

Before giving a string theory and a field theory proof of this
duality, we will first give an example. Consider a $U(1)$ theory
with $n$ fundamentals, all with charge one. The dual theory has 
$U(1)^{n-1}$ gauge symmetry and charge matrix
\be
\left(
\begin{array}{ccc}
1\!\!&& \\[-0.1cm]
&\!\ddots\!& \\[-0.1cm]
&&\!\! 1
\end{array} \left| 
\begin{array}{c}
-1 \\[-0.1cm] 
\vdots \\[-0.1cm]
-1 \end{array} \right.
\right)
\ee
and after a change of basis one sees that this is precisely the dual
gauge theory proposed in \cite{si}.

\subsection{String Theory Proof}

In \cite{si},\cite{SS} it was proposed that mirror symmetry in three-dimensional
gauge theories should be a consequence of T-duality of type IIA and IIB string 
theory. In \cite{dhoo} we elaborated on this proposal to provide evidence for the
dualities proposed in that paper. In addition to this, various D-brane techniques
to study three-dimensional gauge symmetries and their mirror symmetry have been
developed \cite{s,dhoo,pz,hw}. In the latter case, matter usually
appears through open strings connecting D-branes, and is therefore charged under
at most two gauge groups. Since we are interested in the case where matter can
be charged under arbitrary many gauge groups, the string theory analysis using
T-duality will be more useful. 

We therefore view the three-dimensional gauge theories as being obtained from
a compactification of the type IIA string on $CY \times S^1_R$. To get a field
theory in three dimensions, we have to take a limit where $R \rightarrow 0$ and
$\alpha' \rightarrow 0$, while keeping $R/\sqrt{\alpha'}$ finite. The type IIA
compactification is T-dual to IIB on $CY \times S^1_{\alpha'/R}$, and the radius
of this circle also shrinks to zero. Under T-duality, the vector and hypermultiplet
moduli spaces are interchanged, which corresponds exactly to
what happens in mirror symmetry in three dimensions. 

In order to apply T-duality, we first have to construct a singular Calabi-Yau 
manifold so that type IIA compactified on it will give rise to a $U(1)^{N_c}$
gauge theory in three dimensions, with matter with charge matrix $({\bf 1}|m)$.
Only the local structure of the singularity is relevant for the field theory
\cite{kkv}. In order to get $N_f$ hypermultiplets coming from wrapping D2-branes
on two-cycles, we need $N_f$ vanishing two-cycles, which we will denote by
$C_i$ and $C_{\alpha}$. Furthermore, in order to get a $U(1)^{N_c}$ gauge
symmetry, we need that these $N_f$ two-cycles satisfy $N_f-N_c$ relations in
homology, so that there are only $N_c$ homologically independent two-cycles.
Since the charges of the hypermultiplets can be read off from the homology
relations, we find that in our case the homology relations have to be
$[C_{\alpha}] = \sum_i m_{i\alpha} [C_i]$. For finite sizes of the two-cycles
we are in the Coulomb branch of the gauge theory. After applying T-duality
we end up in the Higgs branch of the dual theory. It is very difficult to
recognize the gauge field and matter content of a gauge theory in the Higgs 
phase, and therefore it is better to first go to the Higgs phase of the
original theory, and to apply T-duality after that, so as to end up in
the Coulomb branch of the dual theory. The transition of the Coulomb branch
to the Higgs branch is geometrically given by a conifold or extremal
transition \cite{strom,gms,gmv}. 
	In type IIA string compactification,
     it is the process in which two-cycles shrink
     to points which are then deformed to three-cycles
     with finite volume.
Denote by $\tilde{C}_i$ and $\tilde{C}_{\alpha}$
the three cycles obtained after shrinking the corresponding two-cycle
and replacing it by a three-cycle. To find out the homology relations
between these three-cycles, we use the results in \cite{gmv}. Before
shrinking the two-cycles, there are ``magnetic'' four cycles $C_i^{\ast}$
which are dual to $C_i$ in homology. In other words, the intersection numbers
are $<C_i^{\ast} | C_j> = \delta_{ij}$ and 
$<C_i^{\ast} | C_{\alpha} >=m_{i\alpha}$. After the  conifold transition,
these magnetic four-cycles have become four-chain
(denoted again by $C_i^{\ast}$), with boundary given
by 
\be \label{j1}
\partial C_i^{\ast} = \sum_j <C_i^{\ast} | C_j> \tilde{C}_j + 
\sum_{\alpha} <C_i^{\ast} | C_{\alpha} > \tilde{C}_{\alpha}.
\ee
There are
therefore $N_c$ relations between the $N_f$ three-cycles, given by
\be \label{j2}
[\tilde{C}_i] = -\sum_{\alpha} m_{i\alpha} [\tilde{C}_{\alpha}].
\ee
The Calabi-Yau with these vanishing three-cycles describes the Higgs
phase of the original theory if we put the type IIA string on it, and
after T-duality it should describe the Coulomb branch of the dual theory
if we put the type IIB string on it. In type IIB compactifications, 
vanishing three-cycles give rise to matter by wrapping D3-branes on them,
the number of $U(1)$ gauge groups is the number of homologically independent
three-cycles, and the charges can be read off from the homology relations.
By inspection of (\ref{j2}), we see that the dual theory is a $U(1)^{N_f-N_c}$
gauge theory, with charge matrix $({\bf 1}|-m^t)$. 
Notice that the homology relations (\ref{j2}) are such that the conifold
transition is indeed possible \cite{gmv}.
This gives
a string theory proof of the proposed duality. In order to also derive the
mirror map and to find the explicit forms of the metrics on the Coulomb
and Higgs branches of the moduli space, we now turn to a field theory
proof of the duality.

\subsection{Field Theory Proof}

The field theory proof will simply consist of an explicit computation of the
metrics on the Higgs and Coulomb branches of the moduli space of the
$U(1)^{N_c}$ gauge theory. The Coulomb branch of the moduli space is
given a hyperk\"ahler manifold, whose metric can be computed in perturbation
theory. Because the gauge group is Abelian, there can be no monopole
corrections to the metric and perturbation theory should give the full answer.
Each $U(1)$ vector multiplet contains three scalars, whose expectation 
values we denote by $\vec{r}_i$, $i=1, \ldots, N_c$. In addition, there are
$N_c$ angular variables $\phi_i$ which is the expectation value of the
scalar dual to the gauge fields, and which is periodic with period $2\pi$.
Constant shifts of these angular variables $\phi_i$ is a symmetry of
the theory that is unbroken in perturbation theory, and gives rise to $N_c$
triholomorphic $U(1)$ isometries of the Coulomb branch metric. Such
metrics can always be written in the form \cite{liro,pepo}
\be \label{j3}
ds^2 = g_{ij} d\vec{r}_i d\vec{r}_j + (g^{-1})_{ij}
 (d\phi_i + \vec{\omega}_{ik} \cdot d\vec{r}_k)
 (d\phi_j + \vec{\omega}_{jl} \cdot d\vec{r}_l)
\ee
where 
\bea
\vec{\nabla}_i g_{jk} & = & 
\vec{\nabla}_j g_{ik} \nonumber \\{}
\frac{\partial}{\partial r^p_i} \omega^q_{jk} -
\frac{\partial}{\partial r^q_j} \omega^p_{ik} & = &
\epsilon_{pqr} 
\frac{\partial}{\partial r^r_i} g_{jk}. \label{j5}
\eea
Either by comparing to known cases, or by doing a direct one-loop
calculation, one finds that through one loop the metric $g_{ij}$
is given by
\be \label{coulmetr}
g_{ij} = \frac{\delta_{ij}}{e^2} + \frac{\delta_{ij}}{|\vec{r}_i|} 
  + \sum_{\alpha} \frac{m_{i\alpha} m_{j\alpha}}{|\vec{r}_k m_{k\alpha} 
  + \vec{m}_{\alpha}|}
\ee
where $e$ is the bare gauge coupling. Mirror symmetry will 
only hold in the limit where 
$e\rightarrow \infty$. Although we have no rigorous proof, 
it seems plausible that $g_{ij}$
does not receive any further higher loop corrections. 
If there would be a good off-shell
formulation of $N=4$ multiplets in three-dimensions, the first term
of the metric (\ref{j3}) would come from an F-term,
and one could use this to argue in favor of the absence of higher 
loop corrections to $g_{ij}$.  
The absence of higher loop corrections is also supported by the string theory 
considerations
in the previous section, and we will assume this to be the case from now on.

It remains to compute the metric on the Higgs branch, which
is given by an hyperk\"ahler quotient of an
$N_f-N_c$ quaternionic dimensional vector space with the flat metric
by the action of the group $U(1)^{N_c}$. Luckily, the metric on the Higgs
branch is not subject to perturbative or non-perturbative corrections, and
the classical considerations will be exact.
The moment map equations for the hyperk\"ahler quotient are certain quadratic
equations for the expectation values of the scalar fields in the hypermultiplets.
However, since we are dealing with an Abelian hyperk\"ahler quotient, we
can perform a change of variables that linearizes the moment map equations,
see \cite{giry} and section~4.3 of \cite{dhoo}. In this change of variables, we
replace the two complex numbers that are the expectation values of the 
two complex scalars in a hypermultiplet by a three vector and an angular
variable. Denote the vector and angular variable coming from $<Q_i>$ by
$\vec{s}_i,\psi_i$, and those coming from $<Q_{\alpha}>$ by 
$\vec{s}_{\alpha},\psi_{\alpha}$. Then the flat metric for $<Q_i>$ and
$<Q_{\alpha}>$ in terms of the new variables reads
\bea \label{j11}
ds^2 & = & \sum_i \left(\frac{1}{|\vec{s}_i|} d\vec{s}_i d\vec{s}_i + 
              |\vec{s}_i| (d\psi_i + \vec{\omega}_i \cdot d\vec{s}_i)^2 \right)
    \nonumber \\{} & & +
    \sum_{\alpha} \left(\frac{1}{|\vec{s}_{\alpha}|} d\vec{s}_{\alpha} 
              d\vec{s}_{\alpha} + 
              |\vec{s}_{\alpha}| (d\psi_{\alpha} + \vec{\omega}_{\alpha}
               \cdot d\vec{s}_{\alpha})^2 \right),
\eea
where $\vec{\omega}$ is determined by means of (\ref{j5}). The $\vec{s}$ are
clearly the natural variables to compare the Higgs branch metric to the Coulomb
branch metric. The moment map equations in terms of the new variables
read
\be \label{j9}
\vec{\mu}_i \equiv \vec{s}_i + \sum_{\alpha} m_{i\alpha} \vec{s}_{\alpha}
 = \vec{\zeta}_i, \qquad i=1,\dots,N_c
\ee
where the $\vec{\zeta}$ are the Fayet-Iliopoulos parameters. The vector fields that
generate the action of $U(1)^{N_c}$ are
\be V_i = \frac{\partial}{\partial \psi_i} + \sum_{\alpha} m_{i\alpha} 
\frac{\partial}{\partial \psi_{\alpha}}.
\ee
The moment map equations can be used to solve for $\vec{s}_i$ 
in terms of $\vec{s}_{\alpha}$. 
If we substitute this solution in (\ref{j11}), we find the metric on the constrained 
manifold $\vec{\mu}^{-1}(\vec{\zeta})$. Subsequently, we can use the $U(1)^{N_c}$ symmetry
to gauge fix $\psi_i=0$, so that $\vec{s}_{\alpha},\psi_{\alpha}$ are the coordinates
on the hyperk\"ahler quotient ${\cal M}=\vec{\mu}^{-1}(\vec{\zeta})/U(1)^{N_c}$. 
To compute the metric on the hyperk\"ahler quotient, we need to project its tangent 
vectors in the direction perpendicular to the gauge group. For example, 
let us compute the inner product of the vectors 
$V_{\alpha}=\partial/\partial \psi_{\alpha}$ 
and $V_{\beta}=\partial/\partial \psi_{\beta}$ on ${\cal M}$. If $(,)$ denotes the
inner product on $\vec{\mu}^{-1}(\vec{\zeta})$, and $(,)_{\cal M}$ that on
${\cal M}$, then 
\be
(V_{\alpha},V_{\beta})_{\cal M} = (V_{\alpha},V_{\beta}) - (V_{\alpha}, V_i) (N^{-1})^{ij} 
(V_j,V_{\beta})
\ee
where $N_{ij}=(V_i,V_j)$. Explicitly, this becomes
\bea
(V_{\alpha},V_{\beta})_{\cal M} & = & |\vec{s}_{\alpha}| \delta_{\alpha\beta} - 
 |\vec{s}_{\alpha}| m_{i\alpha} (N^{-1})^{ij} |\vec{s}_{\beta}| m_{j\beta} \nonumber \\  
N_{ij} & = & |\vec{s}_i| \delta_{ij} + m_{i\alpha} |\vec{s}_{\alpha}| m_{j\alpha} 
\eea
where $\vec{s}_i$ is expressed in terms of $\vec{s}_{\alpha}$ by means of (\ref{j9}). 
After some algebra, one can show that $(V_{\alpha},V_{\beta})_{\cal M}$ is the
inverse of the matrix $g_{\alpha\beta} = |\vec{s}_{\alpha}|^{-1} \delta_{\alpha\beta} 
 + m_{i\alpha} |\vec{s}_i|^{-1} m_{j\beta}.$ In a similar fashion one can
find all the remaining components of the metric on ${\cal M}$, with as final
result that the metric on the Higgs branch is given by  
\be \label{j15}
ds^2 = g_{\alpha\beta} d\vec{s}_{\alpha} d\vec{s}_{\beta} + (g^{-1})_{\alpha\beta}
(d\psi_{\alpha} + \vec{\omega}_{\alpha\gamma} \cdot d\vec{s}_{\gamma})
(d\psi_{\beta} + \vec{\omega}_{\beta\delta} \cdot d\vec{s}_{\delta})
\ee
with 
\be \label{higgsmetr}
g_{\alpha\beta} = \frac{\delta_{\alpha\beta}}{|\vec{s}_{\alpha}|} + \sum_i 
   \frac{m_{i\alpha} m_{i\beta} }{|\vec{\zeta}_i - m_{i\gamma} \vec{s}_{\gamma} |}.
\ee
By comparing (\ref{coulmetr}) and (\ref{higgsmetr}), we immediately see that the
duality $m \leftrightarrow -m^t$ indeed exchanges the metrics on the Higgs and Coulomb
branches, and that the mirror map is given by (\ref{mirrormap}).

\subsection{General Charges}

The duality we considered so far was restricted to the case where the charge
matrix had the specific form $({\bf 1}|m)$. Here we consider what happens if
the charge matrix has a generic form $(a_{ij}|b_{i\alpha})$, with $a_{ij}$ a
non-degenerate matrix. This situation can always be achieved, if necessary after
a relabeling of the hypermultiplets. In addition, we have the freedom
to choose a different basis for the generators of the $U(1)^{N_c}$ gauge group.
In order to keep the same gauge group, and not a multiple cover of it, different
bases must be related by an $N_c \times N_c$ integer-valued matrix $c_{ij}$
with determinant $\pm 1$. Thus, the theory with charge matrix $(a_{ij}|b_{i\alpha})$ 
is equivalent to the theory with charge matrix $(c_{ik}a_{kj}|c_{ik}b_{k\alpha})$.
If the determinant of $a_{ij}$ is $\pm 1$, we can choose $c=a^{-1}$, and we
are back in the situation we already discussed. It remains therefore to discuss
the case where the determinant of $a$ is not equal to $\pm1$. 

First, we consider what happens to a general metric of the form (\ref{j3}) if
we introduce new variables $\phi_i = c_{ij} \phi'_j$, 
with $c_{ij}$ some non-degenerate
integer-valued matrix. If we at the same time replace 
$\vec{r}_i = \vec{r}'_j (c^{-1})_{ji}$, $g_{ij}=c_{ik} c_{jl} g'_{kl}$
and $\vec{\omega}_{ij}=c_{ik} c_{jl} \vec{\omega}'_{kl}$, the metric
keeps the form (\ref{j3}) in terms of the primed variables, and
(\ref{j5}) remains satisfied. If we assume that the $\phi'_i$ are also
periodic with period $2\pi$, then the metric in the primed variables 
describes a $\Gamma$-fold cover of the metric in the unprimed variables,
where $\Gamma$ is the finite group ${\bf Z}^n/c({\bf Z}^n)$. 

The above remarks are useful when we study the Higgs branch of the theory
with charges $(a_{ij}|b_{i\alpha})$. The moment map equations read
\be
a_{ij} \vec{s}_i + b_{i\alpha} \vec{s}_{\alpha} = \vec{\zeta}_i
\ee
and these can again be used to solve for $\vec{s}_i$ in terms of
$\vec{s}_{\alpha}$. In addition, the $U(1)^{N_c}$ symmetry can be
used to gauge $\psi_i=0$, but in contrast to the previous case, this
does not yet completely fix the $U(1)^{N_c}$ gauge symmetry. A finite
subgroup $\Gamma$ still acts on the $\psi_{\alpha}$, while preserving
$\psi_i=0$. This means that (\ref{j15}), with $\vec{\zeta}_i$
replaced by $(a^{-1})_{ik}\vec{\zeta}_k$ and $m_{i\alpha}$ by
$(a^{-1})_{ik} b_{k\alpha}$, describes in fact a $\Gamma$-fold cover
of the Higgs branch. The finite group $\Gamma$ is given by
\be \label{j17}
\Gamma = \frac{ {\bf Z}^{N_f} \oplus b^t (a^{-1})^t
 ({\bf Z}^{N_c}) }{ {\bf Z}^{N_f} }.
\ee
In order to find the metric on the Higgs branch itself, we have
to mod out by the action of the group $\Gamma$, which is similar
to the situation described above where we replaced $\phi_i =c_{ij}
\phi'_j$ etc. Here, we need to replace $\psi_{\alpha} = 
c_{\alpha\beta} \psi'_{\beta}$, with similar replacements
for the other variables. In order to correctly implement the
action of the group $\Gamma$, the $N_f \times N_f$ matrix
$c_{\alpha\beta}$ must be chosen in such a way that its 
columns form a basis for the lattice ${\bf Z}^{N_f} \oplus b^t 
(a^{-1})^t ({\bf Z}^{N_c}) $. This then finally yields the 
following metric on the Higgs branch, where we dropped the
primes
\be \label{j19}
g_{\alpha\beta} = \sum_{\gamma} \frac{ (c^{-1})_{\alpha\gamma}
           (c^{-1})_{\beta\gamma} }{|\vec{s}_{\delta} c^{-1}_{\delta\gamma}|} 
           + \sum_i \frac{(a^{-1}b(c^{-1})^t)_{i\alpha} 
           (a^{-1}b(c^{-1})^t)_{i\beta} }{|(a^{-1})_{ij} \vec{\zeta}_j - 
           (a^{-1}b(c^{-1})^t)_{i\gamma} \vec{s}_{\gamma} |}.
\ee
It is straightforward to compute the metric on the Coulomb branch 
through one loop in a theory with arbitrary charge matrix, by
simply generalizing (\ref{coulmetr}). Comparing with (\ref{j19})
we then find that the Higgs branch of a $U(1)^{N_c}$ theory with
charge matrix $(a_{ij}|b_{i\alpha})$ is the same as the Coulomb
branch of a $U(1)^{N_f-N_c}$ theory with charge matrix 
$((c^{-1})_{\alpha\beta}| -(c^{-1} b^t (a^{-1})^t)_{\alpha i})$.
Furthermore, the mirror map reads $\vec{m}_i = (a^{-1})_{ij} 
\vec{\zeta}_j$. As an example, on finds that the 
Higgs branch of a $U(1)^2$ theory with charges
\be
\left(
\begin{array}{ccc} 9 & 3 & 13 \\ 5 & 8 & 7 \end{array} \right)
\ee
is the same as the Coulomb branch of a $U(1)$ theory with charges
\be
\left(
\begin{array}{ccc} 57 & -83 & 2 \end{array} \right).
\ee

At this point we have only established one-half of a mirror
symmetry. We still have to show that the Coulomb branch of the
theory with charge matrix $(a_{ij}|b_{i\alpha})$ is equal to
the Higgs branch of the theory with charge matrix 
$((c^{-1})_{\alpha\beta}| -(c^{-1} b^t (a^{-1})^t)_{\alpha i})$.
We will skip the details, but after a careful analysis one
finds that the Coulomb branch of the $(a|b)$
theory is the quotient of the Higgs branch of the 
$(c^{-1}| -c^{-1} b^t (a^{-1})^t)$
theory by a finite group $Z$. This finite group is the subgroup
of $U(1)^{N_c}$ that acts trivially on all hypermultiplets with
charge matrix $(a|b)$. For generic charges $(a|b)$, this subgroup
will be trivial and we have an exact mirror symmetry. For non-trivial
$Z$, there exists no exact mirror symmetry, but only an approximate
mirror symmetry, where one moduli space is a finite cover of the
other. A different way to phrase the condition that $Z$ is trivial
is to demand that the gauge theory has complete Higgsing, because
$Z$ is the discrete subgroup of the gauge group that survives on
the Higgs branch. This is remarkably 
similar to the condition we encountered in
section~3, and may well be a condition that applies to all
dual pairs.

\section{Instanton Corrections from Type IIA String Theory}

In this section we discuss a D-brane wrapping framework to study instanton
corrections to the metric on the vector multiplet
moduli space of $N=4$ supersymmetric gauge theories in three dimensions.
The field theories arise as 
the particle limit of type IIA string 
theory compactified on $M_6 \times S^1$, where $M_6$ is a 
Calabi-Yau 3-fold and the radius $\varepsilon$ of $S^1_{\varepsilon}$ is sent to zero.
The gauge group and matter content of the gauge theories depend on the nature of the
singularity of the CY 3-fold.
We will be mainly interested in the case of a 3-fold $M_6$  constructed as 
a family of $K3$ fibered over a complex curve of genus $g$.
The singularity that we will consider arises when the $K3$ has singularities of the type
$A_k$ at $n$ isolated points which are resolved to $A_{k-1}$ over a generic point. 

The set up is as follows:
Consider M-theory compactified to three dimensions on a Calabi-Yau 4-fold $M_8$ of the type
$M_6\times T^2$, where $M_6$ is a Calabi-Yau 3-fold. This yields an $N=4$ supersymmetric
three-dimensional 
theory. Denote by $\varepsilon^2$ the area of  $T^2$. In the the limit 
$\varepsilon \rightarrow 0$  we get the 
type IIA string compactified on $M_6 \times S^1_{\varepsilon}$
with  radius of $S^1_{\varepsilon}$ sent to zero.

It was argued in \cite{witten} that the instanton  corrections in the three dimensional theory
arise from the 5-branes in M-theory wrapping divisors of the 4-fold.
Two types of divisors are distinguished:\\
\noindent
(a) Vertical divisors: Divisors of the type $D=C\times T^2$ where C is a divisor of $M_6$.\\
\noindent
(b) Horizontal divisors: In our case $M_6$ itself.

In the limit $\varepsilon \rightarrow 0$ only the vertical divisors are relevant.
Moreover, in the particle limit we only probe the local structure of the singularity and
therefore only the exceptional divisors contribute.

In \cite{witten} a necessary condition for a divisor to contribute to the
superpotential of an $N=2$ theory in three dimensions was found, namely
that its arithmetic genus has to be one.
In our case, the divisors contribute to the $\int d^4\theta$ term and therefore
to the metric on the moduli space,
since there are
four fermionic zero modes associated with the breaking of half of the supersymmetry by
the 5-brane.  Also, the condition on the arithmetic genus $\chi$ of the divisor $D$ is modified to
\beq
\chi(D) =0
\stop
\eeq 
The reason for this change compared to \cite{witten} is that
the $U(1)$ charge of $\int d^4\theta$ is zero while that of 
 $\int d^2\theta$ is one.
In fact it is trivial to see that the arithmetic genus of  a divisor 
of the type $D=C\times T^2$
is always zero.
This however is not a sufficient condition, since there are in general other fermionic zero
modes that may cause the contribution to the metric to vanish.
A sufficient condition is 
\beq
 h_{1,0}(C) = h_{2,0}(C)=0
 \comma
 \eeq
 which means that there are
no fermionic zero modes other than those that arise from the breaking of half of the
supersymmetry by the 5-brane.

Let us now apply this framework in order to see when to expect instanton corrections.
Consider the following examples which follow from the analysis of \cite{klemm,kmp,bkkv,kkv}:\\
\noindent
(i) A singularity of the conifold type where $n$ isolated 2-spheres with $n-1$ linear
relations among them
shrink to zero size.
Stated differently we have a singularity of 
$A_{1}$ type at $n$ points over a curve of genus zero $P^1$, which is resolved at a generic
point.
In the particle limit we have an $N=4$ gauge theory in three dimensions with gauge group
$U(1)$ and $n$ hypermultiplets in the fundamental representation.
Being isolated singularities, the resolution does not lead
to exceptional divisors that can contribute. Thus there are no instanton corrections
in this case. This is compatible with field theory, since there are no monopoles in an
Abelian gauge theory.\\
\noindent
(ii) A singularity of the type
$A_{k-1}$ over a curve of genus zero $P^1$. 
In the particle limit we have an
$N=4$ gauge theory in three dimensions with gauge group
$SU(k)$ and no matter.
There are $k-1$ exceptional divisors of the form $P^1$ over $P^1$. 
These are complex surfaces
$C_i, i=1,...,k-1$ which satisfy  $h_{1,0}(C_i) = h_{2,0}(C_i)=0$. 
Thus we are guaranteed to have instanton
corrections in this case.
 This is compatible with what we expect from field theory. 
 In this case the instanton corrections
 are essential to make the metric on the vector multiplet moduli space positive
 definite.\\
\noindent
(iii) A singularity of type
$A_{k-1}$ over a curve of genus $g$, $\Sigma_g$. 
In the particle limit we have an $N=4$ gauge theory in three dimensions with gauge group
$SU(k)$ and $g$ adjoints.
There are $k-1$ exceptional divisors of the form $P^1$ over $\Sigma_g$. 
These  complex surfaces
do not satisfy  $h_{1,0}(C_i)=0$. Thus we expect no instanton
corrections.
Indeed in the presence of  massless adjoints 
we do not expect instanton corrections 
(nor higher than one loop corrections) \cite{dhoo}.\\
\noindent
(iv) A singularity of type $A_{k}$ at $n$ points which is resolved to 
$A_{k-1}$ over a generic point of a curve of genus zero $P^1$.
In the particle limit we have an $N=4$ gauge theory in three dimensions with gauge group
$U(k)$ and $n$ hypermultiplets in the fundamental representation.
There are $k-1$ exceptional divisors $C_i$ of the form $P^1$ over $P^1$
which satisfy  $h_{1,0}(C_i) = h_{2,0}(C_i)=0$.
Thus we expect instanton
corrections.
We considered a similar example 
in section 2, where we showed that there are open D-string
instantons contributing to the worldvolume field theory on the Dirichlet 3-branes.
\\
\noindent
(v) A singularity of type $A_{k}$ at $n$ points which is resolved to 
$A_{k-1}$ over a generic point of a curve of genus $g$ ,$\Sigma_g$.
In the particle limit we have an $N=4$ gauge theory in three dimensions with gauge group
$U(k)$, $g$ adjoints and $n$ fundamentals.
There are $k-1$ exceptional divisors $C_i$ of the form $P^1$ over $\Sigma_g$. 
These  complex surfaces
do not satisfy  $h_{1,0}(C_i)=0$. Thus we expect no instanton
corrections in this case .
We considered a similar example corresponding
to $g=1$ in section 2, where we showed that there are no open D-string
instantons contributing to the worldvolume field theory on the Dirichlet 3-branes.

In \cite{witten} the instanton contributions to the superpotential of the three dimensional
$N=2$ theories were computed in several cases using the wrapping of the 5-branes.
It would be interesting to do an analogous  computation of the instanton corrections
to the metric on the vector multiplet moduli space using the above framework.

\section*{Acknowledgments}
We would like to thank 
  A.~Hanany, K.~Intriligator, S.~Kachru, H.~Nakajima, M.~Ro\v{c}ek, 
  C.~Schweigert and N.~Seiberg for discussions. 
  This work is supported in part by 
NSF grant PHY-951497 and DOE grant DE-AC03-76SF00098. JdB is a fellow of
the Miller Institute for Basic Research in Science. 
Z. Yin is supported in part by a Graduate Research Fellowship of
the U.S. Department of Education.

\end{document}